\documentclass[12pt]{article}
\pdfoutput=1
\usepackage{jheppub}

\usepackage{slashed}
\usepackage{color, verbatim}
\usepackage{latexsym}
\usepackage{epsfig}
\usepackage{amsmath,accents}

\usepackage{amssymb}
\usepackage{graphicx}
\usepackage{bm}
\usepackage{stackrel}

\usepackage[font={small}]{caption}

\usepackage{hyperref}
\usepackage{epstopdf}
\definecolor{myred}{rgb}{0.7, 0, 0}
\definecolor{myblue}{rgb}{0, 0, 0.7}
\definecolor{mygreen}{rgb}{0.04, 0.7, 0.5}
\hypersetup{colorlinks,citecolor=myred,linkcolor=myblue,urlcolor=myblue,linktocpage=true}

\makeatletter

\@addtoreset{equation}{section}
\makeatother

\newcommand{\be}{\begin{equation}}
\newcommand{\ee}{\end{equation}}
\newcommand{\bea}{\begin{eqnarray}}
\newcommand{\eea}{\end{eqnarray}}
\newcommand{\tr}{\operatorname{tr}}
\newcommand{\diag}{\operatorname{diag}}

\newcommand{\h}{\mathfrak h}

\newcommand{\A}{\mathcal A}
\newcommand{\BB}{\mathcal B}

\newcommand{\TeV}{{\textrm{TeV}}}

\newcommand{\UV}{{\textrm{UV}}}

\newcommand{\Imaginary}{{\textrm{Im}}}
\newcommand{\PP}{{\mathcal{P}}}

\newcommand{\IImm}{{\textrm{Im}}}

\begin{document}


\begin{center}


\begin{center}

\vspace{.5cm}

{\Large\sc
The Continuum Linear Dilaton\footnote{Contribution to the special volume in memoriam of Prof.~Martinus Veltman}
\vspace{0.3cm}
}\\

\end{center}

\vspace{1.cm}

\textbf{
Eugenio Meg\'ias$^{\,a}$, Mariano Quir\'os$^{\,b}$
}\\

\vspace{.1cm}
${}^a\!\!$ {\em {Departamento de F\'{\i}sica At\'omica, Molecular y Nuclear and \\ Instituto Carlos I de F\'{\i}sica Te\'orica y Computacional, Universidad de Granada,\\ Avenida de Fuente Nueva s/n,  18071 Granada, Spain}}

${}^b\!\!$ {\em {Institut de F\'{\i}sica d'Altes Energies (IFAE),\\ The Barcelona Institute of  Science and Technology (BIST),\\ Campus UAB, 08193 Bellaterra (Barcelona), Spain
}}

\end{center}

\vspace{0.8cm}

\centerline{\bf Abstract}
\vspace{2 mm} 

\begin{quote}\small
  Continuum spectra can be a way out to alleviate the tension generated by the elusiveness of narrow resonances of new physics in direct experimental searches. 
Motivated by the latter, we consider the linear dilaton model with a continuum spectrum of KK modes. It is provided by a critical exponential bulk potential for the scalar field stabilizing the distance, between the UV boundary at $y=0$ and a naked (good) singularity at $y=y_s$,  in proper coordinates, which corresponds in conformal coordinates to $z_s\to\infty$. The cutoff $M_s$ in this theory is an intermediate scale $M_s\simeq 10^{-5}M_{\rm Pl}$ and the warped factor solves the hierarchy between $M_s$ and the TeV, while the hierarchy between $M_{\rm Pl}$ and $M_s$ has to be solved by a (Little) String Theory with coupling $g_s\simeq 10^{-5}$. The Standard Model is localized on a 4D IR brane. The graviton and radion Green's and spectral functions have a continuum of states with a TeV mass gap, and isolated poles consisting on the 4D graviton and the light radion/dilaton. We construct the effective field theory below the mass gap where the continua of KK modes are integrated out, generating a set of dimension eight operators which contribute to low energy electroweak precision observables, and high energy violation of unitarity in vector boson scattering processes. The radion mass depends on the stabilizing UV brane potential and its wave function is localized toward the IR which enhances its coupling with the SM fields.  
\end{quote}

\vfill

\newpage

\tableofcontents

\newpage

\section{Introduction}
\label{sec:introduction}
No clear deviation has been found, so far, at present (LHC, \dots) and past (Tevatron, LEP/SLC, \dots) particle physics experiments, from the predictions of the Standard Model (SM) of electroweak and strong interactions. However, given a number of observational facts which cannot be coped by the SM (dark matter and dark energy, the baryon asymmetry of the universe,...), and some theoretical drawbacks as, among others, its sensitivity to the ultraviolet (UV) scale (a.k.a. hierarchy problem) it is generally believed that the SM is an effective theory, and that some UV completion, with beyond the SM (BSM) physics, is needed. This fact has motivated a plethora of BSM models aiming to UV completing the SM, thus solving some of the above issues. In fact, Martinus Veltman was one of the pioneers to recognize the hierarchy problem and the need for a UV completion of the SM, in a seminal paper published in 1981 by Acta Physica Polonica B~\cite{Veltman:1980mj}.

One of the most successful BSM models was proposed in 1999 by Lisa Randall and Raman Sundrum~\cite{Randall:1999ee}, where the hierarchy between the four-dimensional (4D) Planck scale $M_{\rm Pl}$ and the TeV scale is provided by a warped fifth dimension, in a five-dimensional (5D) space with a non factorizable metric and two branes, the UV brane and the infrared (IR) brane. This theory predicts the existence of a discrete spectrum made out of towers of Kaluza-Klein (KK) states, with masses in the TeV range, associated with the SM fields, as e.g.~the graviton. However the elusiveness of narrow resonances in direct searches~\cite{Sirunyan:2018ryr,Aaboud:2019roo} led people to imagine different solutions to the hierarchy problem (leaving aside the possibility of superheavy KK modes~\cite{Megias:2020vek,Megias:2021rgh}), either with broad resonances~\cite{Escribano:2021jne} or even a continuum of resonances heavier than a mass gap~\cite{Csaki:2018kxb,Megias:2019vdb}, evading direct searches and thus detectable only by indirect measurements.   

In theories with a warped extra dimension, the brane distance is stabilized by means of a bulk scalar field $\bar\phi$ with brane potentials (the Goldberger-Wise mechanism~\cite{Goldberger:1999uk}), and different theories are classified by the behavior of the bulk scalar field in the limit $\bar\phi\to\infty$~\cite{Cabrer:2009we}. In fact, in the absence of the IR boundary, it is found in Ref.~\cite{Cabrer:2009we} that there is a critical behavior for the spectrum to be a continuum with a mass gap, when the bulk potential behaves as $V(\bar\phi)\propto e^{2\bar\phi}$ in the limit $\bar\phi\to\infty$.  If the bulk potential goes faster than $e^{2\bar\phi}$ there is a discrete spectrum, and if it goes slower there is a continuum spectrum without any mass gap. For this critical behavior of the potential the stabilizing field $\bar\phi$ behaves linearly in conformally flat coordinates $z$ near the asymptotic limit. This linear behavior was obtained as the 5D effective theory of a class of type II strings, known as Little String Theory (LST)~\cite{Aharony:1999ks}, in the decoupling limit of very small string coupling. In this paper we will then study 5D warped theories with a bulk potential which is $V(\bar\phi) \propto e^{2\bar\phi}$, for all values of the field $\bar\phi$, thus providing a warped realization of the so-called linear dilaton models. 

We have found two classes of theories, depending on the sign of the metric slope with respect to the conformal coordinate. \textit{i)} For the case of negative slope~\cite{Antoniadis:2011qw,Cox:2012ee,Giudice:2017fmj}, gravity decouples in the limit of $z\to\infty$ so that an IR brane is compelling. The size of the bulk cutoff is the TeV so that the SM fields should be localized on the UV brane and the whole hierarchy problem has to be solved by the LST, with a string scale at the TeV and a coupling as tiny as $\sim 10^{-15}$. In the presence of the brane the graviton and radion KK modes are discrete with a mass gap, and separated by $\sim 30$ GeV. \textit{ii)} For the case of positive slope, in the limit $z\to\infty$ gravity is correctly described for a cutoff at an intermediate scale $\sim 10^{-5} M_{\rm Pl}$, so that the string scale has to be fixed by the LST at that intermediate scale and the string coupling is small $\sim 10^{-5}$, but larger than in the case of negative slope. The SM has to be localized on the IR brane, which is not a boundary of the space, and the spectrum for the graviton and radion is continuum with a gap related to the metric slope. The distance between the UV boundary and the IR brane is stabilized by a Goldberger-Wise mechanism. 

The discrete spectrum of Kaluza Klein gravitons in Randall-Sundrum theories was considered in Refs.~\cite{Davoudiasl:1999jd,Davoudiasl:2001uj,Fitzpatrick:2007qr,Agashe:2007zd,Antipin:2007pi,Dillon:2016fgw}.
In this paper we have studied the linear dilaton theory, with positive slope metric in conformally flat coordinates and continuum graviton and radion spectra. The contents of the paper are as follows. In Sec.~\ref{sec:general} the gravitational background is analyzed in detail, with the discussion of linear dilaton models with different sign slopes, and their connection with Little String Theory. The Green's functions and spectral functions for the graviton are studied in Sec.~\ref{sec:graviton}. The graviton Green's functions contain a massless isolated pole, corresponding to the 4D graviton and a continuum of KK modes with a TeV mass gap. The coupling of the graviton continuum with the SM fields is studied in Sec.~\ref{subsec:coupling_to_matter}, where a class of dimension eight operators is obtained after integrating out the continuum of KK modes in the effective theory. The radion Green's and spectral functions are studied in Sec.~\ref{sec:radion}. The radion Green's function has a continuum of resonances and an isolated pole in the first Riemann sheet corresponding to a mass, below the continuum mass gap, which depends on the UV stabilizing brane potential. Integrating out the continuum leaves an effective theory with dimension eight operators similar to the case of the graviton KK modes. Our conclusions are drawn in Sec.~\ref{sec:conclusions}.

\section{The gravitational background}
\label{sec:general}

We consider a slice of 5D space-time between a brane at the value $y = y_0=0$ in proper coordinates, the UV boundary brane, and a (possible) admissible singularity~\cite{Gubser:2000nd} placed at $y = y_s$, a value which has to be determined dynamically. In addition, we will introduce an IR brane, at $y = y_1<y_s$, responsible for electroweak breaking, where we will assume the SM sector to be localized. 

The 5D action of the model, with metric defined by 
\begin{eqnarray}
ds^2 &=&g_{MN}dx^M dx^N\equiv e^{-2A(y)} \eta_{\mu\nu} dx^\mu dx^\nu - dy^2 \,,  \label{eq:metric}  
\end{eqnarray}
including the stabilizing bulk scalar $\phi(x,y)$, with mass dimension $3/2$, reads as
\begin{eqnarray}
S &=& \int d^5x \sqrt{|\det g_{MN}|} \left[ -\frac{1}{2\kappa^2} R + \frac{1}{2} g^{MN}(\partial_M \phi)(\partial_N \phi) - V(\phi) \right]\nonumber \\
&-& \sum_{\alpha} \int_{B_\alpha} d^4x \sqrt{|\det \bar g_{\mu\nu}|} \lambda_\alpha(\phi)  
 -\frac{1}{\kappa^2} 
 \int_{B_0} d^4x \sqrt{|\det \bar g_{\mu\nu}|} K_0  \,, \label{eq:action}
\end{eqnarray}
where $\kappa^2=1/(2M^3_5)$, $M_5$ being the 5D Planck scale, $V(\phi)$ and $\lambda_\alpha(\phi)$ are the bulk and brane potentials of the scalar field $\phi$, and the index $\alpha=0$, 1 refer to the UV and IR branes, respectively. The IR brane is responsible for the generation of the IR scale $\sim \TeV$, and contains the brane Higgs potential which spontaneously breaks the electroweak symmetry, thus solving the hierarchy problem between $M_5$ and the TeV scale for the considered value of $A(y_1)$, as we will see. 
In Eq.~(\ref{eq:action}) the 4D induced metric is $ \bar g_{\mu\nu}=e^{-2A(y)}\eta_{\mu\nu}$, where the Minkowski metric is given by $\eta_{\mu\nu} =\diag(1,-1,-1,-1)$. 
 The last term in Eq.~(\ref{eq:action}) is the usual Gibbons-Hawking-York boundary term~\cite{York:1972sj,Gibbons:1976ue}, where $K_{0}$ is the extrinsic UV curvature. In terms of the metric of Eq.~(\ref{eq:metric}) the extrinsic curvature term reads as~\cite{Megias:2018sxv} $K_{0} =  -4 A^\prime(y_{0})$. Note that the extrinsic curvature at the singularity is canceled by the action of the determinant.

The equations of motion (EoM) read then as
\begin{eqnarray}
&&A^{\prime\prime}
= \frac{\kappa^2}{3} \phi^{\prime \, 2} + \frac{\kappa^2}{3} \sum_\alpha \lambda_\alpha(\phi) \delta(y - y_\alpha)  \,, \label{eq:eom1}\\
&&A^{\prime\, 2} 
= -\frac{\kappa^2}{6} V(\phi) + \frac{\kappa^2}{12} \phi^{\prime\, 2} \,,  \label{eq:eom2}\\
&&\phi^{\prime\prime} - 4 A^\prime \phi^\prime = V^\prime(\phi) + \sum_\alpha \lambda_\alpha^\prime(\phi) \delta(y - y_\alpha)  \,, \label{eq:eom3}
\end{eqnarray}
where the prime symbol $(\,{}^\prime\,)$ will hereafter stand for the derivative of a function with respect to its argument.
The EoM in the bulk can also be written in terms of the superpotential $W(\phi)$ as~\cite{DeWolfe:1999cp}
\begin{equation}
\phi^\prime = \frac{1}{2} \frac{\partial W}{\partial \phi} \,, \qquad A^\prime = \frac{\kappa^2}{6} W \,, \label{eq:phiA}
\end{equation}
and
\begin{equation}
V(\phi) = \frac{1}{8} \left( \frac{\partial W}{\partial \phi} \right)^2 - \frac{\kappa^2}{6} W^2(\phi) \,. \label{eq:V}
\end{equation}

The brane potential terms in the EoM are responsible for boundary and jumping conditions for the fields in the branes. In particular, by integrating the equations in a neighborhood of each brane, and using Eq.~(\ref{eq:phiA}), we get on the UV boundary
\be
W(\phi(y_0))=\lambda(\phi(y_0)) \,,\qquad W^\prime(\phi(y_0))=\lambda^\prime(\phi(y_0)) \,, \label{eq:BCUV}
\ee
where the $\mathbb Z_2$ orbifold conditions have been used. On the other hand, on the IR brane we have to impose continuity conditions for $W(\phi)$ and $W^\prime(\phi)$, i.e.~
\be
\Delta W(\phi(y_1))=0 \,, \qquad \Delta W^\prime(\phi(y_1))=0 \,, 
\label{eq:BCIR}
\ee
where $\Delta X$ is the jump when crossing the brane. Simple brane potentials satisfying the boundary (\ref{eq:BCUV}) and jumping (\ref{eq:BCIR}) conditions, and fixing dynamically the values of $\phi$ at the branes, i.e.~$v_\alpha\equiv \phi(y_\alpha)$, are given by
\be
\lambda_0(\phi)=W(\phi)+\frac{1}{2}\gamma_0(\phi-v_0)^2\,, \qquad  \lambda_1(\phi)=\frac{1}{2}\gamma_1(\phi-v_1)^2  \,.
\label{eq:lambda01}
\ee

Integrating the gravitational 5D Lagrangian by parts in the bulk, and after using the background EoM, one can see that there are contributions to the potential localized on the boundaries as $\mp e^{-4A(y_\alpha)}W$~\cite{Gubser:1999vj}, where the $\mp$ sign corresponds to the boundaries $y_\alpha=(y_0,\,y_s)$. While this contribution vanishes on the singularity at $y=y_s$, it gives a contribution to the UV boundary such that the effective UV brane potential is
\be
U_0(\phi)=\lambda_0(\phi)-W(\phi)=\frac{1}{2}\gamma_0(\phi-v_0)^2  \,, \label{eq:U0}
\ee 
which is dynamically minimized for $\phi=v_0$. Moreover, on the IR brane there is not such boundary contribution and there the effective potential is $U_1(\phi)=\lambda_1(\phi)$ which is minimized for the value of $\phi=v_1$.

For convenience we will define the dimensionless field $\bar\phi\equiv\kappa\phi/\sqrt{3}$. The properties of the 5D theory depend on the superpotential behavior in the limit $\bar\phi\to\infty$~\cite{Cabrer:2009we}. In particular when the asymptotic superpotential behavior is exponential $e^{\nu\bar\phi}$, for $\nu<1$ the spectrum is continuous without mass gap, for $\nu>1$ there is a mass gap and a discrete spectrum, and for the critical value $\nu_c=1$ the spectrum is continuous with a mass gap.

We will hereby consider the critical case $\nu_c$ where the superpotential  and bulk potential are 
\be
W(\bar\phi) = \frac{6k}{\kappa^2}e^{\bar\phi},\quad V(\bar\phi) = -\frac{9k^2}{2\kappa^2}e^{2\bar\phi} \,,
\label{eq:W_lineardilaton}
\ee
where $k\lesssim M_5$ is the parameter which determines the 5D curvature. The model defined by the superpotential (\ref{eq:W_lineardilaton}) has a singularity at a finite value of the proper coordinate $y=y_s$, as in soft wall models. It leads to a gapped continuum spectrum, and the hierarchy problem is, more conventionally, solved in the same way as in RS theories, with fundamental scales $M_5$ and $k$, and a derived TeV scale after warping. The solution for the background in proper coordinates is~\footnote{The solution of the EoM, Eq.~(\ref{eq:phiA}), leads in fact to
$A(y) = \bar\phi(y) + c$,
where $c$ is a constant that can be fixed by choosing $A(0) = 0$ so that  $c = -\bar \phi(0) = -\bar v_0$.
}
\begin{equation}
\bar\phi(y)=-\log[k(y_s-y)],\quad A(y)=-\log(1-y/y_s) \,. \label{eq:phiy_Ay}
\end{equation}

We will also consider, as in general soft wall models, two branes, at $y=0$ (the UV boundary) and $y=y_1$ (the IR or Higgs brane) where we assumed the Standard Model, and in particular the Higgs, to be located, such that the values of the IR brane location $y_1$ and the singularity $y_s$ will be dynamically determined by the brane potentials ·$\lambda_\alpha(\phi)$, fixing the field $\bar\phi$ at the values $\bar v_0\equiv\kappa v_0/\sqrt{3} $ and $\bar v_1\equiv \kappa v_1/\sqrt{3}$ in the UV and IR branes, respectively, such that
\begin{equation}
ky_s=e^{-\bar v_0},\quad ky_1=e^{-\bar v_0}-e^{-\bar v_1} \,. \label{eq:kys_ky1}
\end{equation}
Notice that the first expression demands that $k y_s > 0$. Moreover, the solution of the hierarchy problem between $M_5$ and the TeV scale is achieved for a given value of the warp factor at the IR brane $A(y_1)\equiv A_1$, which imposes the relation
\begin{equation}
\bar v_1 - \bar v_0 = A_1 \,.
\end{equation}

As we will see in Secs.~\ref{sec:graviton} and \ref{sec:radion}, the squared mass gap of the continuum graviton and radion spectrum is then given by 
\begin{equation}
m_g^2 = \frac{9}{4} \rho^2 \,,
\end{equation}
with
\be
\rho = \pm 1/y_s  \,.
\label{eq:rho}
\ee
Using that $dz = \pm e^{A(y)} dy$, where $\pm$ corresponds to the sign of $\rho$, the relation between conformally flat and proper coordinates in the model of Eq.~(\ref{eq:W_lineardilaton}) turns out to be 
\begin{equation}
\rho \cdot ( z - z_0)  = - \log( 1 - y/y_s) \,, \label{eq:z_y}
\end{equation}
and then the background in these coordinates is given by
\begin{equation}
\bar\phi(z) =  A(z) + \bar v_0  \,, \qquad A(z) = \rho \cdot (z - z_0) \,, \label{eq:phiALD}
\end{equation} 
where we fix $z_0\equiv 1/k$. Therefore in conformally flat coordinates the dilaton is linear and the corresponding model is dubbed as linear dilaton model (LDM).

According with Eq.~(\ref{eq:rho}) the  parameter $\rho$ can have both signs, and accordingly two classes of theories are implemented.

\subsection{The $\rho<0$ case: the discrete LDM}
\label{subsec:LST}

Let us now consider the solution with a negative sign of the parameter $\rho$ in (\ref{eq:rho})~\cite{Antoniadis:2011qw,Cox:2012ee}: $\rho=-1/y_s$. Now we will fix the values of the scalar field in the branes as $\bar v_0\simeq 0$ and $\bar v_1\simeq-|A_1|$ and then the relation between $\rho$ and $k$ is given by
\be
k=|\rho|e^{-\bar v_0}\simeq |\rho|\, .
\ee
In this case we will consider as fundamental interval $z_0<z<z_1<\infty$, with two branes located on them. The interval $[z_0,z_1]$ is then mapped into the interval $[-|y_1|,0]$ in proper coordinates and the metric and background field profiles are given by
\begin{align}
A(z)&=-|\rho|(z-z_0) \,, \quad A(y)=-\log(1+|\rho y|)  \,, \\
\bar\phi(z)&=A(z)+\bar v_0 \,,\qquad\, \bar\phi(y)=A(y)+\bar v_0  \,,
\end{align}
so that $A(y)<0$, $A_1\equiv A(y_1)=\bar v_1-\bar v_0<0$. Then
\be
|\rho y_1|=e^{|A_1|}-1\simeq e^{|A_1|},\quad |\rho|(z_1-z_0)=|A_1| \,,
\ee
and the location of the IR brane is dynamically fixed by $\bar v_0$ and $\bar v_1$.

The relationship with the 4D Planck scale is here given by
\be
\kappa^2 M_{\rm Pl}^2=\int_0^{z_1}e^{-3A}dz=\frac{1}{3|\rho|}\left( e^{3|A_1|}-1  \right)\Rightarrow\ M_{\rm Pl} \simeq
\left( \frac{2 M_5^3}{3|\rho|} \right)^{1/2} e^{3|A_1|/2} \,.
\label{eq:Mplnegative}
\ee
The hierarchy problem is solved at the string level, as $M_5\sim |\rho|\sim$ TeV the warped factor in this theory is required to be
$|A_1|\simeq 23 $, which leads to a value of $|y_1| \simeq 10^{-7}$cm, small as compared to that of the ADD model, but greater (by a factor $e^{|A_1|})$ than the one for the $\rho>0$ case. 

The spectrum of this theory is discrete with the lightest KK mode mass being~$\mathcal O(\rho)$. Let us notice that this theory does not admit a continuum spectrum, as sending $z_1\to\infty$ leads to $M_{\rm Pl}\to \infty$, which means that gravity is decoupled. As the aim of this paper is considering continuum spectra for linear dilaton models, this class of models with $\rho<0$ will not be considered here~\footnote{Detailed phenomenological studies of this model have been done in Refs.~\cite{Antoniadis:2011qw,Cox:2012ee,Giudice:2017fmj}.}. 
In this theory the hierarchy problem has to be entirely solved by the string theory which should set the cutoff scale in the 5D theory at the TeV. Therefore here the SM is located in the UV brane.
Notice that in this setup $M_5\simeq \rho\simeq$ TeV are fundamental scales, while the 4D Planck scale is a derived scale.

\subsection{The $\rho>0$ case: the continuum LDM}
\label{subsec:continuum_model}

In this paper we will consider the $\rho>0$ case in (\ref{eq:rho}), and the IR brane location in conformally flat coordinates $z_1$ is dynamically determined, as well as the value of $y_1$, by the IR fixing of the dilaton at the value $\bar v_1$. We will here fix $\bar v_1=0$ and $\bar v_0=-A_1<0$. Then, one finds from Eq.~(\ref{eq:kys_ky1}) that $k y_s = e^{A_1}$ and $k(y_s - y_1) = 1$. 
The hierarchy problem is then solved by fixing $A_1$ such that $\rho=\mathcal O(\textrm{TeV})$, as given by Eq.~(\ref{eq:rho}) for $k\lesssim M_5$ with
\be
\rho= k \, e^{-A_1} \,.
\ee

The value of $M_5$ is determined by the relation of $M_5$ and $k$ with the 4D Planck scale $M_{\rm Pl}$ given by
\be
\kappa^2 M_{\rm Pl}^2=\int_0^{y_s}e^{-2A}dy \quad \Rightarrow \quad M_5=\left(\frac{3}{2}\rho M_{\rm Pl}^2  \right)^{1/3} \,, \label{eq:Mpl}
 \ee
which yields, for $\rho=\mathcal O(\textrm{TeV})$, $M_5\simeq 10^{13}\textrm{ GeV}\simeq10^{-5}M_{\rm Pl}$, and correspondingly a warp factor $A_1\simeq 23$. In conformal coordinates the location of the branes in units of $\rho$ are at $\rho z_0 = e^{-A_1}$ and $\rho z_1 = A_1 + e^{-A_1}\simeq A_1$, while the singularity is located at infinity, $z_s\to\infty$. The length of the fundamental region is $y_1\simeq y_s\simeq 10^{-17}$ cm, much larger than the corresponding one in the RS model $\sim 10^{-31}$ cm, but still much smaller than that in ADD theories~\cite{ArkaniHamed:1998rs}~$\sim 10^{-2}$ cm (for the case of two extra dimensions) where only gravity propagates in the bulk.

In this case, we can consider two kinds of fundamental intervals for our theory:

\begin{itemize}
\item
When the fundamental interval is $[0,y_1]$, i.e.~$[z_0,z_1]$ in conformal coordinates, the theory spectrum is discrete, the first mode mass being $\mathcal O(\rho)$.
\item
When the fundamental interval is $[0,y_s]$, i.e.~$[z_0,\infty)$ in conformal coordinates (as we will consider in this paper), this theory predicts a continuum spectrum with an $\mathcal O(\rho)$ gap: the \textit{continuum linear dilaton model} (CLDM).
\end{itemize}

Let us notice that in this theory the warp factor solves the hierarchy problem between the intermediate scale $M_5$ and the TeV scale. This means that the UV completion of this theory should be a string theory with the string mass at the intermediate scale, $M_s\simeq M_5$, thus solving the hierarchy problem between the Planck scale and $M_5$. This class of models does not support gauge bosons propagating in the bulk of the extra dimension, so that we will consider the whole SM localized at the IR brane. Notice that in this setup, as in RS models, $M_5$ is a fundamental scale, while the TeV scale and the 4D Planck mass are derived from the theory warp factor.

\subsection{Connection with Little String Theory}

As we have seen in the previous section the hierarchy problem between the 4D Planck scale, $M_{\rm Pl}$, and the 5D Planck scale, $M_5\simeq 10^{13}$ GeV, has to be solved by a string theory with string scale at the intermediate value $M_s\simeq M_5$. As the relation between $M_s$ and $M_{\rm Pl}$ in string theory is given by
\be
M_{\rm Pl}^2=\frac{1}{g_s^2}M_s^8 V_6 \,,
\ee  
where $g_s$ is the string coupling and $V_6$ the volume of the compactified dimensions, imposing $M_s\ll M_{\rm Pl}$ requires, either 
$V_6\gg \ell_s^6$, with $\ell_s=1/M_s$, or $g_s\ll 1$. The second possibility, i.e.~$V_6 \sim l_s^6$ and  $g_s\sim M_s/M_{\rm Pl}\ll1$ can be realized in type II string theories, where the size of the gauge coupling is unrelated to $g_s$, but fixed by the geometry (radii) of the compact dimensions where gauge interactions propagate. In the limit $g_s\to 0$ a type II string theory, dubbed Little String Theory~\footnote{For a review see Ref.~\cite{Aharony:1999ks}.}, was constructed where non-abelian gauge interactions are localized on a stack of (Neveu-Schwarz) NS5-branes, a 6D space with 1+3 flat dimensions and two extra longitudinal dimensions compactified on a torus $T^2$ with size $\sim\ell_s^2$~\footnote{The four extra transverse dimensions are compactified in a manifold, and we will assume all compact dimensions have a size $\sim \ell_s$.}. In the gravity decoupling limit $g_s\to 0$ the NS5-branes give rise to the 6D LST, which is strongly coupled and seems to have no Lagrangian description. By holography one can relate the 6D strongly coupled theory in the absence of gravity to a 7D theory with gravity weakly coupled with a linear dilaton. 

Upon compactification of the two extra dimensions in $T^2$ we obtain a 5D theory with weakly interacting gravity and a linear dilaton, as that studied in the previous section. Here there is a fundamental difference between the two previous theories with $\rho<0$ and $\rho>0$. 
\begin{itemize}
\item
In the case of $\rho<0$, by making the extra dimension infinite, i.e.~$z_1\to \infty$, one gets from Eq.~(\ref{eq:Mplnegative}) that $M_{\rm Pl}\to\infty$ and so gravity is exactly decoupled as in the LST with $g_s\to 0$. When introducing the IR brane at a finite value of $z_1$ we recover the correct value of $M_{\rm Pl}$ for $|A_1|\simeq 23$, in which case $g_s=M_s/M_{\rm Pl}\sim 10^{-15}$. As $M_5\sim$ TeV, the 5D theory does not need to solve any hierarchy problem, and the SM is localized in the UV brane. Reproducing the 4D Planck scale means that the fundamental interval is finite and the graviton KK spectrum is discrete, with the mass of the first KK mode being $\mathcal O(3\rho/2)$.
\item
However, in the case of $\rho>0$, and even considering the infinite interval $[z_0,\infty]$, gravity is never decoupled, as one can see from Eq.~(\ref{eq:Mpl}), and the correct value of $M_{\rm Pl}$ is recovered for $g_s\simeq 10^{-5}$. In this case the 5D theory has to solve the hierarchy problem between $M_5$ and the TeV, and one has to introduce the IR (or SM) brane where the SM fields are localized. The distance between the UV and IR branes is precisely dictated by Eq.~(\ref{eq:Mpl}). Here we always have two options: 
\begin{itemize}
\item
A fundamental finite interval $[z_0,z_1]$ in which case there is a discrete KK spectrum for the graviton, with the mass of the first KK mode being $\mathcal O(3\rho/2)$. The discrete mass spectrum is given by
\be
\frac{m_n^2}{\rho^2}=\frac{9}{4}+\frac{\pi^2 n^2}{A^2(z_1)}\,,  \quad (n\in \mathbb Z)
\label{eq:deltam}
\ee
where the squared mass spacing $\Delta m_n^2$ is governed by the value of the warp factor $A_1 \equiv A(z_1)$ at the IR boundary, and decreases for increasing values of $A_1$.
\item
A fundamental infinite interval $[z_0,\infty]$, in which case there is a gap equal to $3\rho/2$, followed by a continuum spectrum. It can be understood from Eq.~(\ref{eq:deltam}) when we take the limit $A(z_1)\to\infty$ so that the squared mass spacing goes to zero. This is the model we will consider in this work.
\end{itemize}
\end{itemize}

In both cases the string theory has to solve (part of) the hierarchy problem and explain the smallness of the coupling $g_s$. Some mechanisms have been proposed in Ref.~\cite{Antoniadis:2001sw}, which induce potentials whose minima should fix the value of $g_s$. Of course, the larger the value of $g_s$ to be fixed, the more easily these mechanisms are satisfied. For (Coleman-Weinberg) potentials which depend logarithmically on $g_s$, as those triggered by an anomalous $U(1)$ with a gauge mass proportional to $\sqrt{g_s}M_s$, where the coefficients ($a$ and $b$) are generated through string loop corrections, as e.g.~$V=g_s^2(a+b\log g_s)M_s^4$, the minimum is at $g_s\simeq e^{-a/b}$ so that, there should be some hierarchy between the coefficients $a$ and $b$, i.e.~$a/b\simeq $ 11 (35) for the case $\rho>0$ $(\rho<0)$.

\section{The graviton}
\label{sec:graviton}

The graviton is a transverse traceless fluctuation of the metric of the form
\begin{equation}
ds^2 = e^{-2A(y)} \left[\eta_{\mu\nu} +2\kappa h_{\mu\nu}(x,y)\right] dx^\mu dx^\nu - dy^2 \,,
\end{equation}
where $h_\mu^\mu = \partial_\mu h^{\mu\nu} = 0$. The Lagrangian is given by
\be
\mathcal L=-\frac{1}{2}\int_0^{y_s} dy e^{-2A}\left[\partial_\rho h_{\mu\nu}\partial^\rho h^{\mu\nu}+e^{-2A}h'_{\mu\nu}h^{\prime \mu\nu}
\right] \,,
\label{eq:grav-lagrangian}
\ee
and we will use the ansatz $h_{\mu\nu}(x,y) =\h(y) h_{\mu\nu}(x)$ from where the EoM can be written as
\be
e^{2A(y)} (e^{-4A(y)}\h^\prime(y))^\prime+p^2\h(y)=0 \,. \label{eq:eomgraviton1}
\ee

In conformal coordinates, cf.~Eq.~(\ref{eq:z_y}), and after rescaling the field by $\h(z) = e^{3 A(z)/2} \tilde\h(z)$, the equation of motion for the fluctuation can be written in the Schr\"odinger like form~\cite{Cabrer:2009we} as
\begin{equation}
-\tilde\h^{\prime\prime}(z) + V_{\h}(z) {\tilde\h}(z) = p^2 {\tilde\h}(z) \,,
\end{equation}
where the potential is given by
\begin{equation}
V_\h(z) = \frac{9}{4} {A^\prime}^2(z) - \frac{3}{2} {A^{\prime\prime}}(z)  \,. \label{eq:Vh}
\end{equation}
An explicit evaluation of this potential with the expression of $A(z)$ given by Eq.~(\ref{eq:phiALD}) leads to a constant value $V_\h(z) = m_g^2$, where $m_g = 3\rho/2$ is the mass gap for the graviton, typical of a continuum of states with a mass gap.

The interaction with matter is found as
\be
\mathcal L_{5D}=-\frac{1}{\sqrt{2}M_5^{3/2}} h^{\mu\nu}(x,y)T_{\mu\nu}(x,y) \,,
\ee
where the energy-momentum tensor is defined as
\be
T_{\mu\nu}=\frac{2}{\sqrt{-g}}\frac{\delta(\sqrt{-g}\mathcal L)}{\delta g^{\mu\nu}} \,,
\ee
and so it is symmetric by definition.

\subsection{Green's functions for the graviton}
\label{subsec:GF_graviton}

The Green's function for $h_{\mu\nu}(x,y)$ in the transverse, traceless gauge, is given by~\cite{Hagiwara:2008jb}
\be
D_{\mu\nu,\rho\sigma}=\frac{1}{2}\left(\eta_{\mu\rho}\eta_{\nu\sigma}+\eta_{\mu\sigma}\eta_{\nu\rho}-\frac{2}{3}\eta_{\mu\nu}\eta_{\rho\sigma} \right) 
G_\h(y,y';p) \,,
\ee
where we are using, for the 5D Green's function $G_\h(y,y';p)$, a mixed representation where the 4D coordinates $x^\mu$ have been Fourier transformed into 4D momenta $p^\mu$. After fixing the value of $y^\prime$, one can see that the Green's function obeys the same EoM as the field $\h(y)$, except for an inhomogeneous Dirac delta term, i.e.
\begin{equation}
p^2 G_\h(y,y^\prime) + e^{2A(y)} \frac{d}{dy}\left( e^{-4A(y)} G_\h^\prime(y,y^\prime) \right) = e^{2A(y^\prime)} \cdot \delta(y-y^\prime) \,,  \label{eq:Ghy}
\end{equation}
where the prime indicates derivative with respect to the variable~$y$ and, for simplicity, we are omitting the $p$ dependence from the argument of the Green's function. After substituting the explicit expression for $A(y)$, we find that the equation writes as
\begin{equation}
G_\h^{\prime\prime}(y,y')  - \frac{4}{y_s - y} G_\h^{\prime}(y,y') + \frac{1}{(y_s - y)^2}\left( \frac{p}{\rho} \right)^2  G_\h(y,y') =  e^{4 A(y^\prime)}  \delta(y-y^\prime)  \,,  \label{eq:eom_graviton3}
\end{equation}
whose general solution is
\begin{equation}
G_\h(y,y^\prime;p) =  \left\{ 
\begin{array}{cc}
C^{I}_1 \cdot (y_s - y)^{\frac{3}{2} \Delta_\h^-} + C^{I}_2 \cdot (y_s - y)^{\frac{3}{2} \Delta_\h^+ }& \quad  y < y^\prime < y_s  \\
C^{II}_1 \cdot (y_s - y)^{\frac{3}{2} \Delta_\h^-} + C^{II}_2 \cdot (y_s - y)^{\frac{3}{2} \Delta_\h^+} &  \quad y^\prime < y < y_s
\end{array} \,, \right. \label{eq:Gh}
\end{equation}
where we have defined
\begin{equation}
\Delta_\h^\pm = \pm \delta_\h - 1  \,, \qquad   \delta_\h = \sqrt{1-(4/9) \cdot p^2/\rho^2 } \,. \label{eq:Delta_h}
\end{equation}

The Green's function $G_\h(y,y')$ is subject to the following boundary and matching conditions
\begin{align}
\begin{split}
& \hspace{0.46cm} G_\h^\prime(0,y') = 0 \,, \qquad \hspace{0.07cm} \Delta G_\h(y^\prime,y^\prime) = 0 \,, \qquad \Delta G_\h^\prime(y^\prime,y^\prime) = e^{4A(y^\prime)} \,, \\
& \Delta G_\h(y_1,y^\prime) = 0 \,, \qquad \Delta G_\h^\prime(y_1,y^\prime)  = 0 \,.  \label{eq:Gh_bc}
\end{split}
\end{align}
In addition, we should impose regularity in the IR, i.e.~we consider $C_1^{II} = 0$. Then, all the integration constants are fixed.

After conveniently defining the variables
\begin{equation}
y_\downarrow = \min(y,y^\prime) \,, \qquad y_\uparrow = \max(y,y^\prime) \,,
\end{equation}
and implementing the boundary and matching conditions in the general solution of Eq.~(\ref{eq:Gh}), one finds~\footnote{Note that when $y^\prime < y_1$ we could split the domain $y^\prime < y < y_s$ in Eq.~(\ref{eq:Gh}) into two domains: $y^\prime < y \le y_1$ and $y_1 < y < y_s$. However, in that case the jumping conditions in the IR brane, i.e.~continuity of $G_{\h}(y,y')$ and $G_{\h}^\prime(y,y')$ in $y = y_1$, demand that the term $(y_s - y)^{\frac{3}{2} \Delta_\h^-}$ is also absent in  $y^\prime < y \le y_1$, so that the solution would be identical as the one presented in Eq.~(\ref{eq:Gh_graviton}). The solution would also be identical to this formula in the case $y_1 < y^\prime$.} 
\begin{eqnarray}
\hspace{-0.5cm} G_\h(y,y^\prime) = \frac{1}{3\rho} \frac{1}{\delta_\h} (1 - \bar y_\uparrow)^{\frac{3}{2}\Delta_\h^+} \left(  - (1 - \bar y_\downarrow)^{\frac{3}{2}\Delta_\h^-} + \frac{\Delta_\h^-}{\Delta_\h^+} (1 - \bar y_\downarrow)^{\frac{3}{2} \Delta_\h^+}  \right) \,,\label{eq:Gh_graviton}  
\end{eqnarray}
where we are using dimensionless coordinates $\bar y\equiv \rho y$. In particular, in the limit $y \to y_0$ one finds
\begin{eqnarray}
G_\h(y_0,y^\prime) &=& -\frac{2}{3} \frac{1}{\rho} \frac{1}{\Delta_\h^+} (1 -\bar y^\prime)^{\frac{3}{2}\Delta_\h^+} \,. \label{eq:Ghz0zp}
\end{eqnarray}

Notice that the Green's function~(\ref{eq:Gh_graviton}) can be expressed as the product of two functions in the form $G_\h(y,y^\prime) = \A(y_\downarrow) \BB(y_\uparrow)$, and this can be written also as
\begin{equation}
\A(y_\downarrow) \BB(y_\uparrow) = \A(y) \BB(y^\prime) \Theta(y^\prime - y) +  \A(y^\prime) \BB(y) \Theta(y - y^\prime)  \,, \label{eq:AB}
\end{equation}
where $\Theta(x)$ is the step function. Then, it is clear that the Green's function is symmetric under the exchange of $y$ and $y^\prime$, i.e.~it fulfills the property
\begin{equation}
G_\h(y,y^\prime) = G_\h(y^\prime,y) \,. \label{eq:GA_sim}
\end{equation}
This property is not obvious from the EoM, Eq.~(\ref{eq:Ghy}). Another property is
\begin{equation}
\Imaginary \left( \A(y) \BB(y^\prime) \right) = \Imaginary \left( \A(y^\prime) \BB(y) \right) \,,  \label{eq:ImAB}
\end{equation}
which follows from the explicit expression of Eq.~(\ref{eq:Gh_graviton}), and taking into account the relation
\begin{equation}
\left( \Delta_\h^{\pm}(p) \right)^\ast = \Delta_\h^{\mp}(p) \,, \qquad (p^2 \ge m_g^2) \,, \label{eq:Deltapm}
\end{equation}
which is valid for time-like momenta $p^2 > 0$. The properties given by Eqs.~(\ref{eq:ImAB}) and (\ref{eq:Deltapm}) will be relevant for the study of the spectral functions in Sec~\ref{subsec:spectral_function}.

Let us consider, in particular, the analytical expressions for the UV-to-UV, UV-to-IR and IR-to-IR Green's functions. There are
\begin{eqnarray}
G_\h(y_0,y_0;p) &=& - \frac{2}{3\rho}  \frac{1}{\Delta_\h^+}  \,, \\
G_\h(y_0,y_1;p) &=& -\frac{2}{3\rho}  \frac{1}{\Delta_\h^+} e^{-\frac{3A_1}{2} \Delta_\h^+}  \,,  \label{eq:Gh_z0z1_asymp}  \\
G_\h(y_1,y_1;p) &=&  \frac{1}{3\rho}  \frac{e^{3A_1
}}{\delta_\h} \left( -1 + \frac{\Delta_\h^-}{\Delta_\h^+} e^{-3A_1\delta_\h}  \right) \,, \label{eq:Gh_z1z1_asymp} 
\end{eqnarray}
respectively. All Green's functions include the zero-mode contributions which behave as
\begin{equation}
G_\h^0 =  \frac{3\rho}{p^2} = \lim_{p \to 0} G_\h(y,y^\prime;p) \,,  \label{eq:Gh0}
\end{equation}
so that we can define Green's functions contributed only by the continuum of KK modes, with the zero-mode contribution subtracted out, as
\be
\mathcal G_\h(y,y')\equiv G_\h(y,y')-G_\h^0 \,.
\label{eq:Gnozeromode}
\ee

\begin{figure}[t]
\centering
\includegraphics[width=4.5cm]{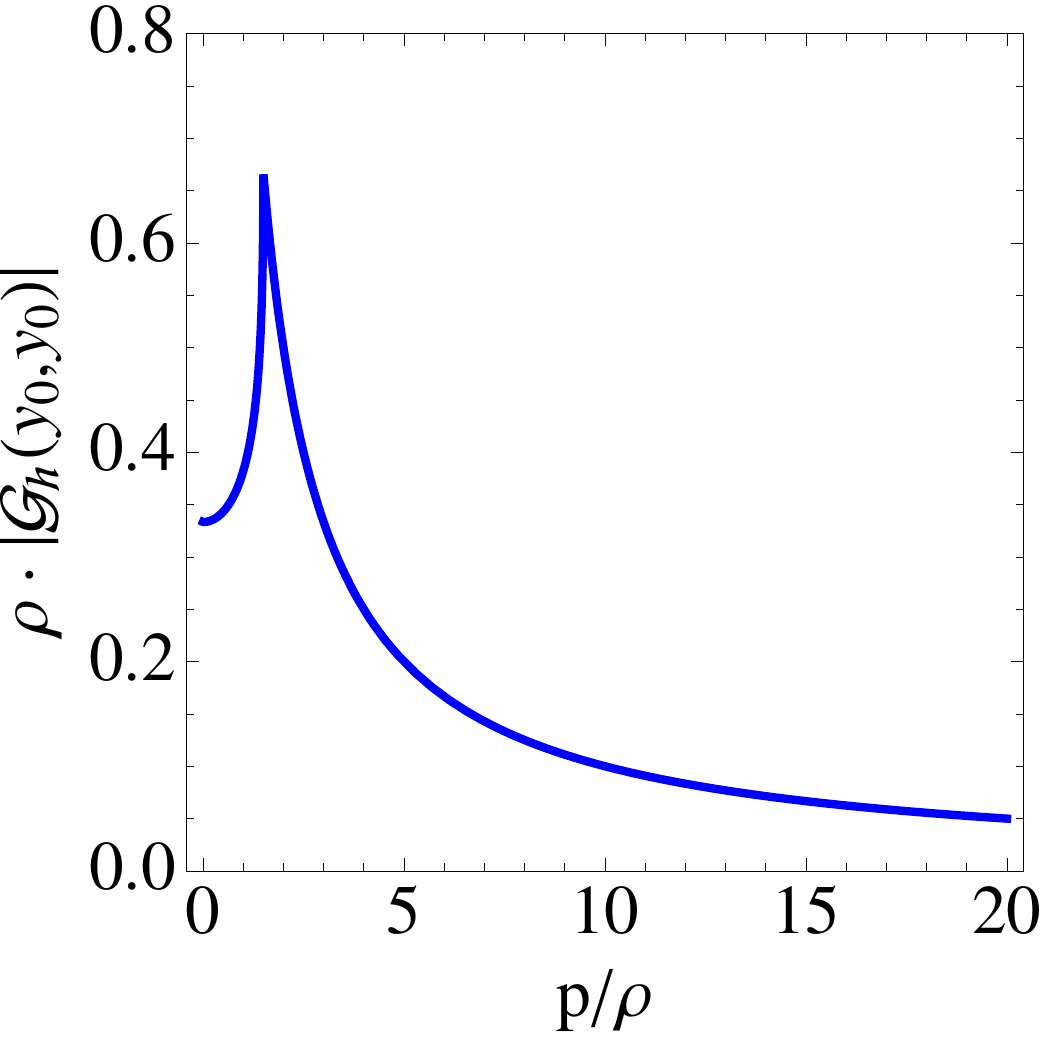} \hspace{0.3cm}
\includegraphics[width=4.5cm]{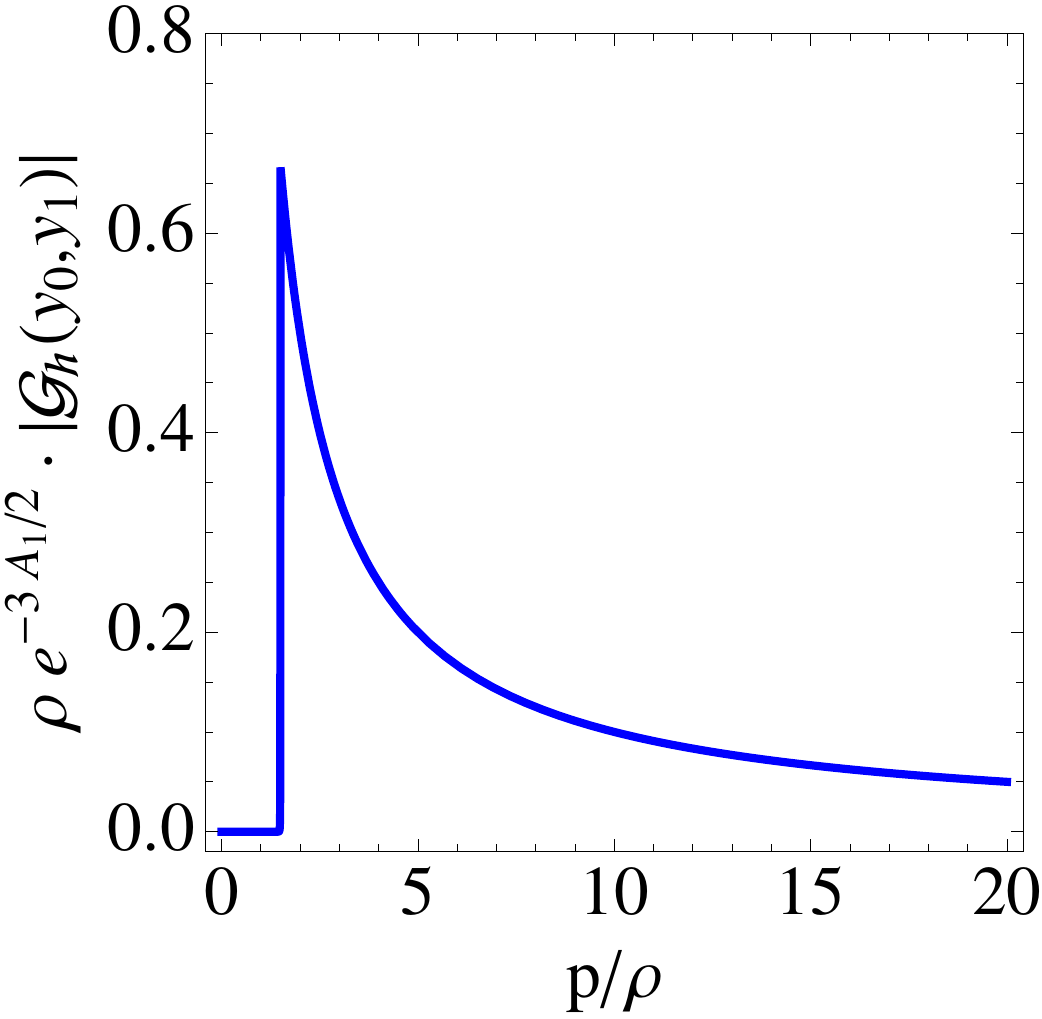} \hspace{0.3cm}
\includegraphics[width=4.7cm]{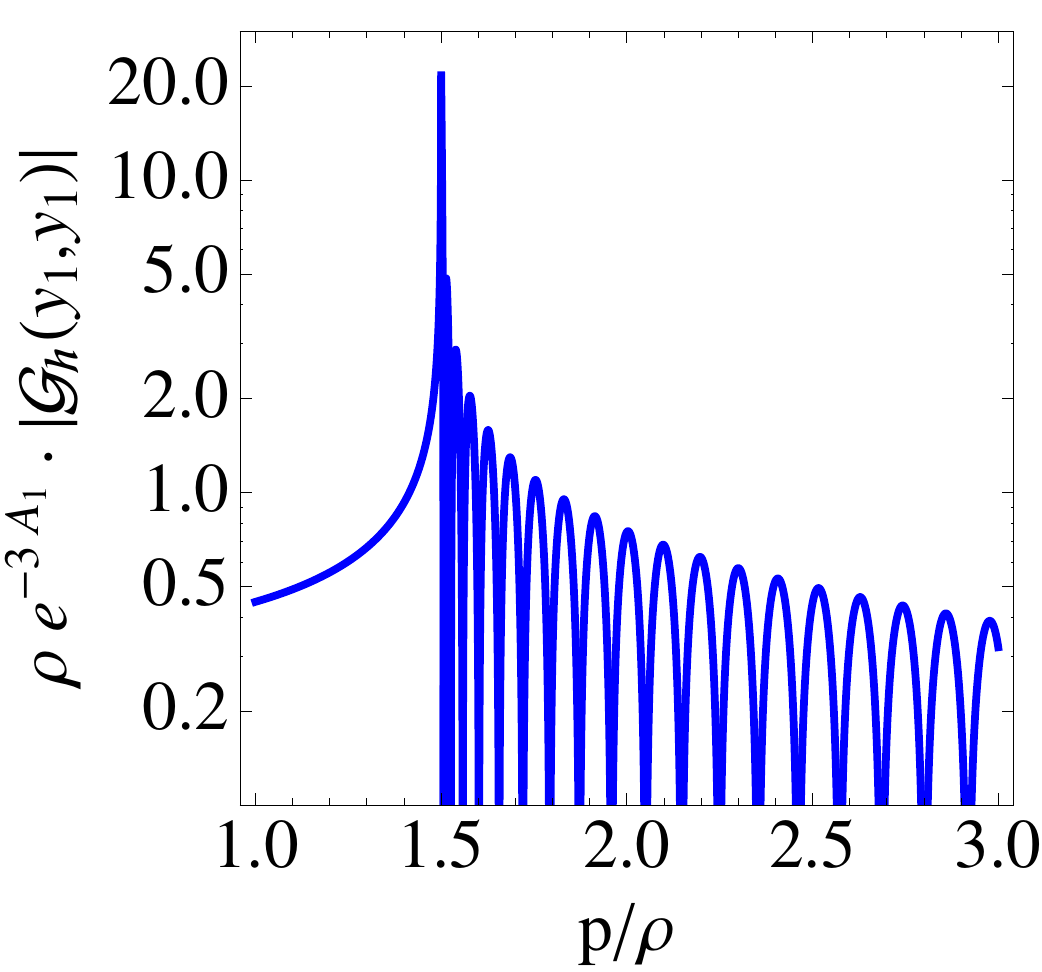}
\caption{\it Plots of $\rho|\mathcal G_\h(y_0,y_0;p)|$ (left panel), $\rho e^{-3A_1/2}|\mathcal G_\h(y_0,y_1;p)|$ (middle panel), and $\rho e^{-3A_1}|\mathcal G_\h(y_1,y_1;p)|$ (right panel) as functions of $p/\rho$. We have used $A_1 = 23$ in all panels and assume time-like momenta $p^2>0$.
}
\label{fig:Ggraviton}
\end{figure}

\begin{figure}[t]
\centering
\includegraphics[width=4.6cm]{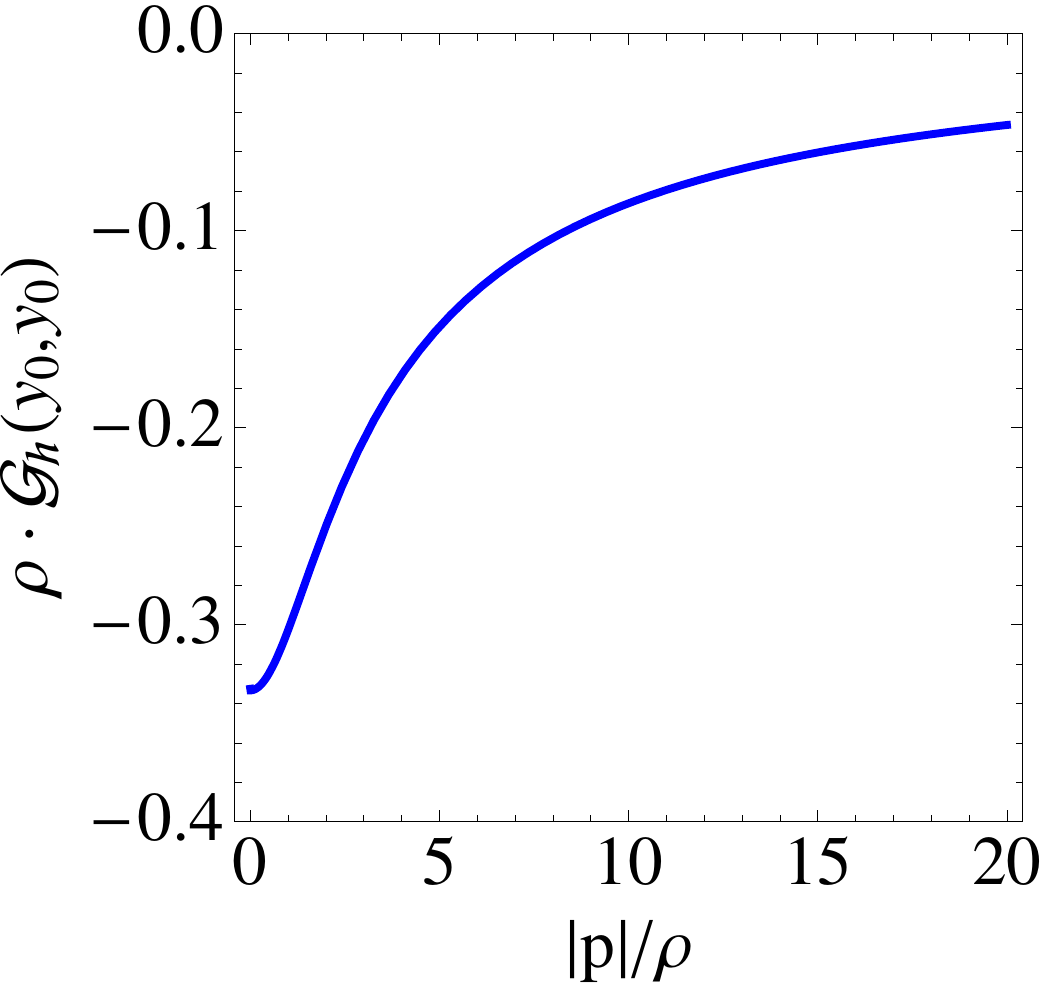}  \hspace{0.3cm}
\includegraphics[width=4.4cm]{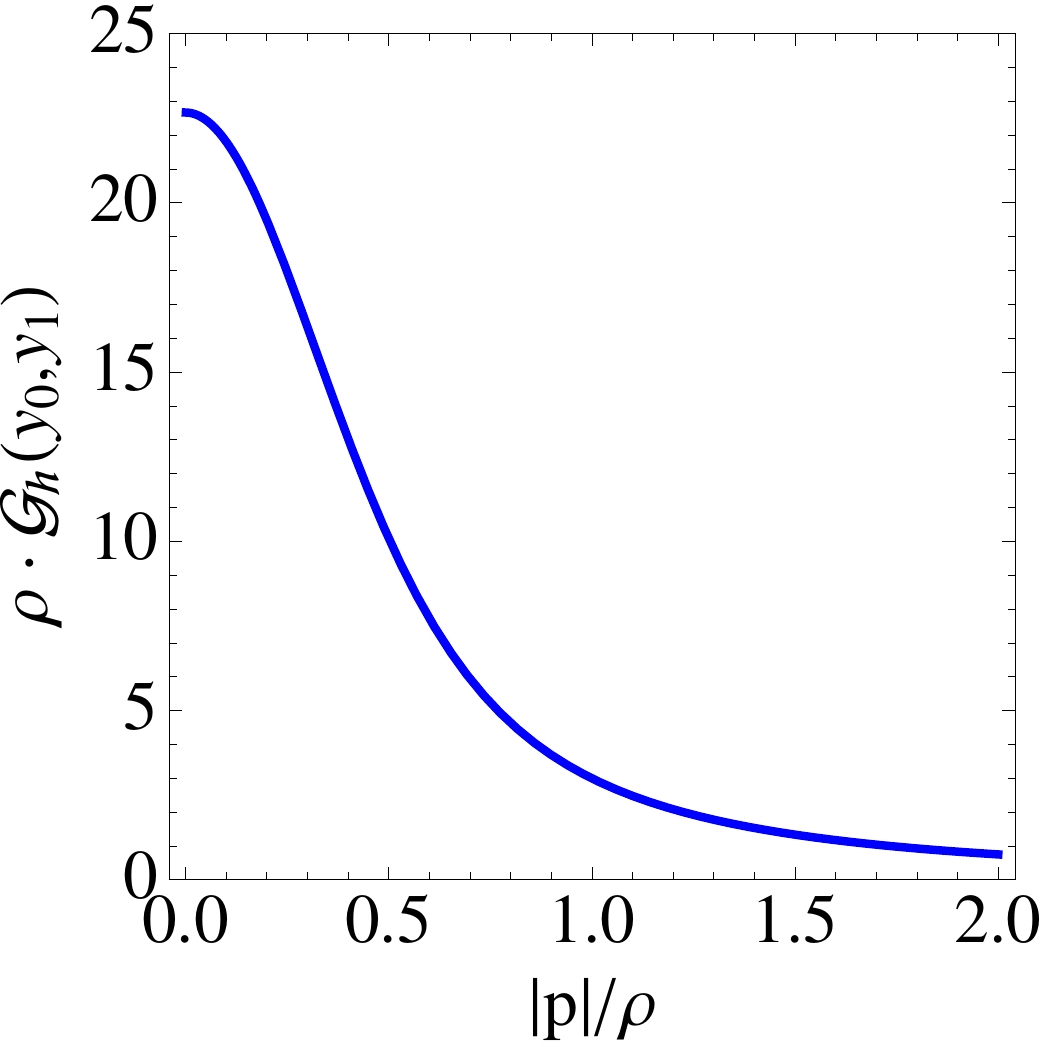}  \hspace{0.3cm}
\includegraphics[width=4.6cm]{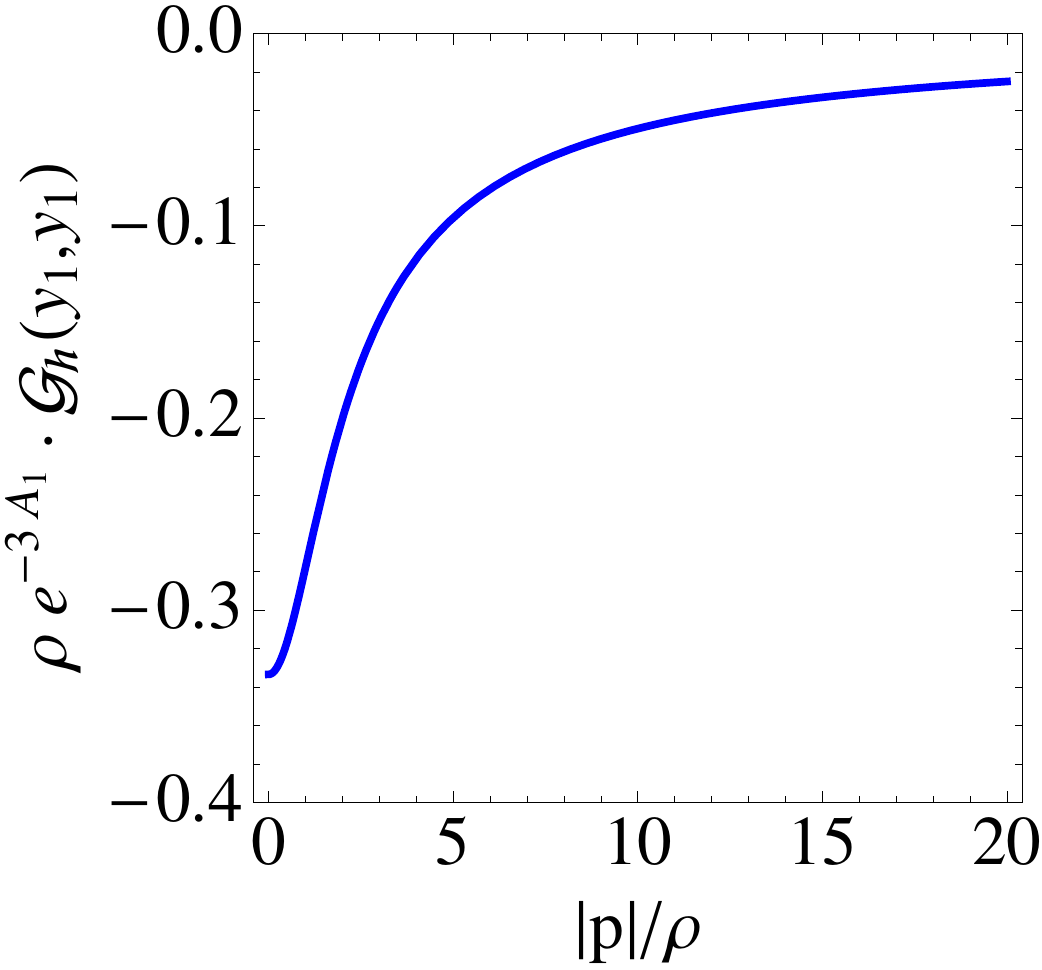}
\caption{\it Plots of $\rho \, \mathcal G_\h(y_0,y_0;|p|)$ (left panel), $\rho \, \mathcal G_\h(y_0,y_1;|p|)$ (middle panel), and $\rho \, e^{-3A_1}\mathcal G_\h(y_1,y_1;|p|)$ (right panel) as functions of $|p|/\rho$. We have used $A_1 = 23$ in all panels and assume space-like momenta $p^2<0$.
}
\label{fig:Ggraviton_pE}
\end{figure}

We plot in Fig.~\ref{fig:Ggraviton}, $|\mathcal G_\h(y_0,y_0;p)|$ (left panel), $|\mathcal G_\h(y_0,y_1;p)|$ (middle panel), and $|\mathcal G_\h(y_1,y_1;p)|$ (right panel), conveniently normalized, as functions of $p/\rho$, for time-like momenta $p^2 > 0$. For space-like momenta $p^2 < 0$ the Green's functions are purely real. We plot in Fig.~\ref{fig:Ggraviton_pE} the Green's functions $\mathcal G_\h(y_0,y_0;|p|)$, $\mathcal G_\h(y_0,y_1;|p|)$, and $\mathcal G_\h(y_1,y_1;|p|)$ as functions of $|p|/\rho$, for space-like momenta $p^2 < 0$.

It is interesting, for further purposes as we will see, to explicitly provide the limits $p \ll \rho$ of the Green's functions. This leads to the following Taylor series expansions
\begin{eqnarray}
G_\h^{-1}(y_0,y_0)  &\stackrel[ p \ll \rho ]{\simeq}{}&  \frac{1}{3} \frac{p^2}{\rho} + \frac{1}{27} \frac{p^4}{\rho^3} + \mathcal O(p^6)  \,, \\
G_\h^{-1}(y_0,y_1) &\stackrel[ p \ll \rho ]{\simeq}{}& \frac{1}{3} \frac{p^2}{\rho} + \frac{1}{27} \left[ 1 - 3 A_1\right] \frac{p^4}{\rho^3} + \mathcal O(p^6)  \,,  \label{eq:Gh_z0z1_asymp}  \\
G_\h^{-1}(y_1,y_1) &\stackrel[ p \ll \rho ]{\simeq}{}&  \frac{1}{3} \frac{p^2}{\rho} + \frac{1}{27} \left[ e^{3A_1} - 6 A_1 \right] \frac{p^4}{\rho^3} + \mathcal O(p^6)  \,. \label{eq:Gh_z1z1_asymp} 
\end{eqnarray}
The asymptotic expansions in the regime $\rho \ll p$, for time-like momenta $p^2 > 0$, are
\begin{eqnarray}
&& G_\h^{-1}(y_0,y_0) \stackrel[ \rho \ll p ]{\simeq}{} i p + \frac{3}{2}\rho   \,, \\
&& G_\h^{-1}(y_0,y_1) \stackrel[ \rho \ll p ]{\simeq}{} i  e^{-A_1(i \frac{p}{\rho} + \frac{3}{2})} p   \,, \\
&& G_\h^{-1}(y_1,y_1) \stackrel[ \rho \ll p ]{\simeq}{} 2i \left[ 1 +    e^{i 2A_1\frac{p}{\rho} } \right]^{-1} e^{-3A_1} p  \,.
\end{eqnarray}
It is now obvious that for time-like momenta, for which $p=i|p|$, $$G_\h(y_0,y_1)\stackrel[|p|\to\infty]{\simeq}{} e^{-A_1|p|/\rho}$$ goes to zero exponentially (see also Ref.~\cite{Costantino:2020vdu}), while both $G_\h(y_0,y_0)$ and $G_\h(y_1,y_1)$ behave has $\sim 1/|p|$.

\subsection{Spectral functions}
\label{subsec:spectral_function}

\begin{figure}[t]
\centering
\includegraphics[width=4.6cm]{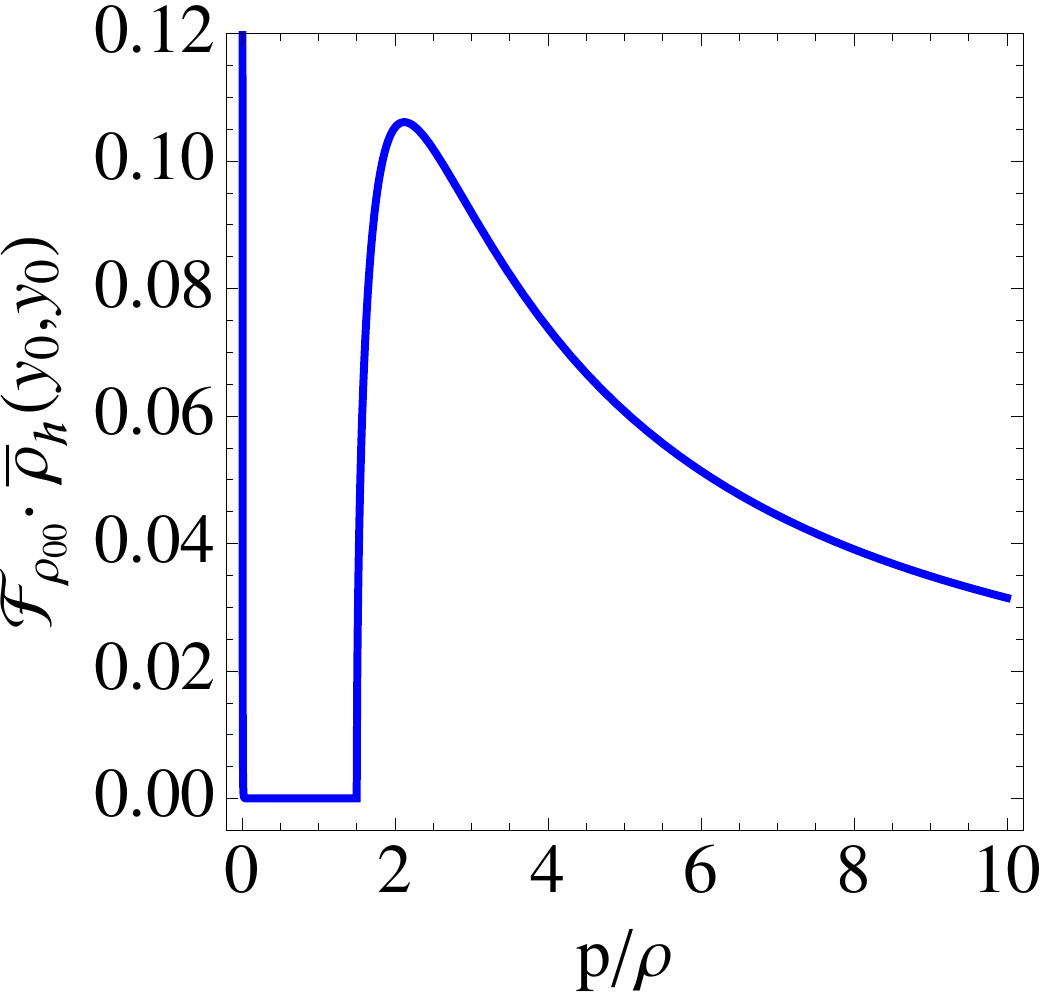}   \hspace{0.3cm}
\includegraphics[width=4.6cm]{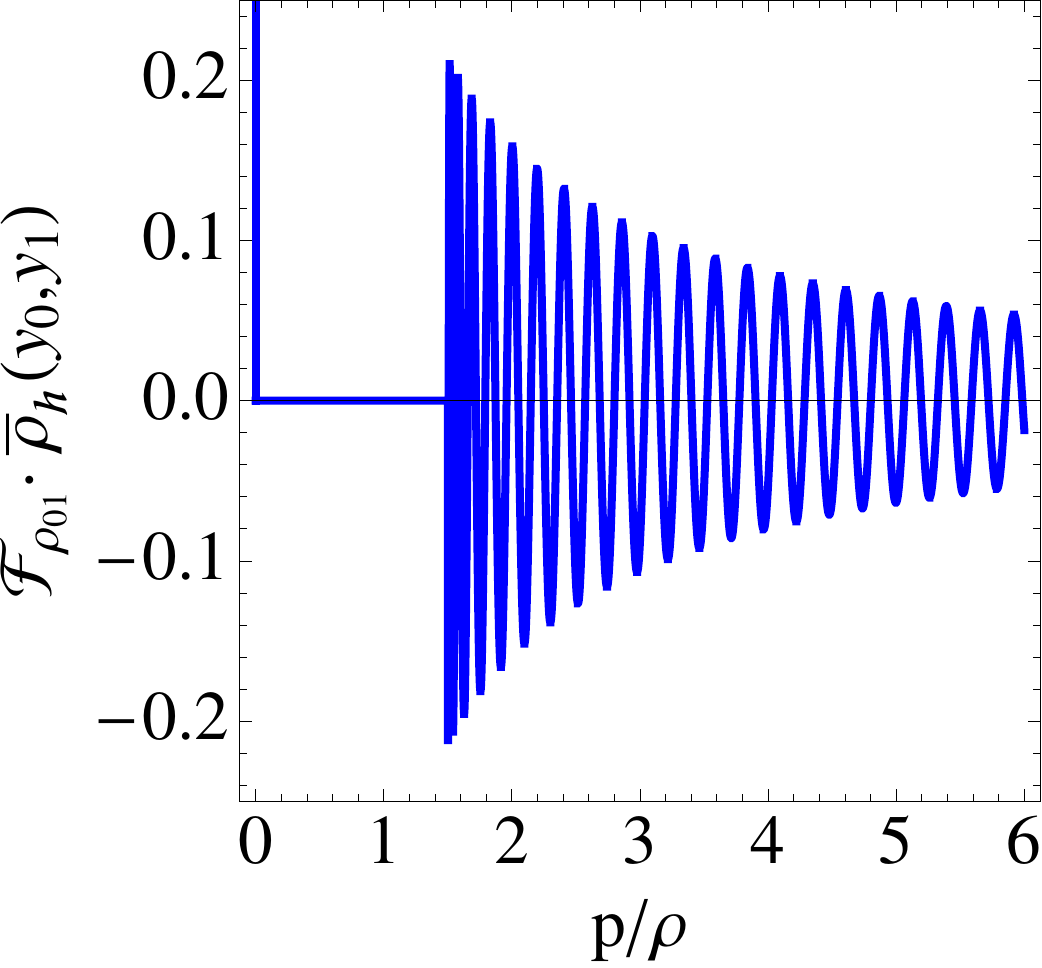}   \hspace{0.3cm}
\includegraphics[width=4.2cm]{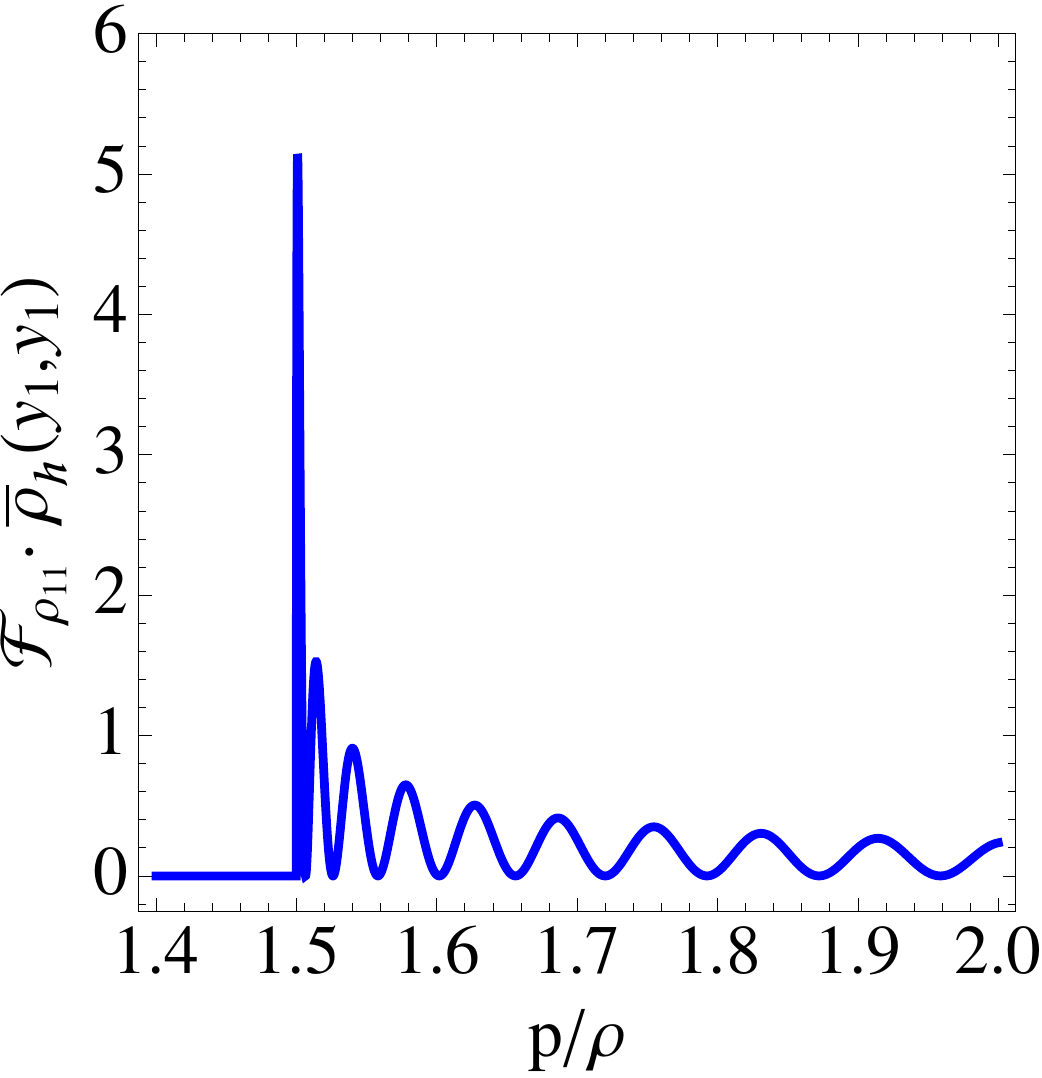} 
\caption{\it Scale invariant spectral functions $\mathcal F_{00} \cdot \bar\rho_\h(y_0,y_0;p)$ (left panel), $\mathcal F_{01} \cdot \bar\rho_\h(y_0,y_1;p)$ (middle panel) and $\mathcal F_{11} \cdot \bar\rho_\h(y_1,y_1;p)$ (right panel) as a function of $p/\rho$, for a continuum graviton. We have used $A_1 = 23$ in all panels and assume time-like momenta $p^2>0$.
}
\label{fig:spectral_graviton}
\end{figure}

In this section we find it convenient to work in a basis with flat extra dimensional coordinate $y$, i.e. with wave function $\bar h_{\mu\nu}(x,y)$ as
\be
\bar h_{\mu\nu}(x,y)=e^{-A(y)}h_{\mu\nu}(x,y) \,,
\ee
with a corresponding Green's function 
\be
\bar G_\h(y,y')=e^{-A(y)}G_\h (y,y')e^{-A(y')} \,.
\ee
For time-like momenta, $p^2>0$, all Green's functions are complex for values of $p > m_g = 3\rho/2$, which is not associated to a particle threshold decay, an intrinsic property of e.g.~unparticle theories~\cite{Georgi:2007ek}. In this way we can define the corresponding spectral functions as
\be
\bar\rho_\h(y,y^\prime;s)= - \frac{1}{\pi} \textrm{Im }\bar G_\h(y,y^\prime;s + i\epsilon) \,, \qquad s \equiv p^2 \,.
\label{eq:spectral}
\ee 
In Fig.~\ref{fig:spectral_graviton} we show $\bar\rho_\h(y_0,y_0;p)$, $\bar\rho_\h(y_0,y_1;p)$ and $\bar\rho_\h(y_1,y_1;p)$, as functions of $p/\rho$  where the prefactors, defined as
\be
\mathcal F_{00} = \rho \,, \qquad  \mathcal F_{01}= \rho\left(\frac{\rho}{k}\right)^{1/2} \,, \qquad \mathcal F_{11} =\rho \left(\frac{\rho}{k}\right) \,,  \label{eq:F_rho}
\ee
make them scale invariant~\cite{Megias:2019vdb}. By using the identity
\begin{equation}
\lim_{\epsilon \to 0^+} \frac{1}{x + i\epsilon} = \PP \frac{1}{x} - i\pi \delta(x) \,, \label{eq:PP}
\end{equation}
one can see that the small $p$ behavior of the Green's functions provided in Sec.~\ref{subsec:GF_graviton} implies the existence of a Dirac delta behavior in the spectral functions at $p=0$,
\begin{equation}
\bar\rho_\h(y,y^\prime;s) = 3\rho e^{-A(y)-A(y')}\, \delta(s) + \cdots \,. \label{eq:rho_delta}
\end{equation}
This delta function appears in all the Green's functions of Fig.~\ref{fig:spectral_graviton}.
Notice that, although the spectral functions $\bar\rho_\h(y_0,y_0)$ and $\bar\rho_\h(y_1,y_1)$ are positive definite, the spectral function propagating from the UV to the IR brane $\bar\rho_\h(y_0,y_1)$ is not. This fact just challenges the physical interpretation of the spectral function in a 4D quantum field theory, which is positive definite by its probabilistic interpretation. 

To understand the positivity of the spectral function in our theory we have to consider, from the 4D point of view, the spectral function $\bar\rho_\h(y,y';s)$ as the matrix element $(y,y')$ of an operator $\hat \rho_\h$, i.e.
\be
(\hat\rho_\h)_{y}^{\,y'}\equiv \bar\rho_\h(y,y';s)
\ee
acting on the infinite dimensional space parametrized by the coordinate $y$. The matrix action on a vector $v_y\equiv v(y)$ is thus represented by the integral, e.g.~$\sum_{y'}(\hat\rho)_{y}^{\,y'}v_{y'}\equiv \int dy' \bar\rho_\h(y,y';s)v(y')$. In parallel with the definition of the operator $\hat \rho_\h$ one can define, from the Green's functions $\bar G_\h(y,y')$ the operator $\hat G_\h$ such that
\begin{equation}
\hat \rho_\h = - \frac{1}{\pi} \Imaginary \; \hat G_\h \,, \qquad \textrm{where} \qquad \Imaginary \; \hat G_\h = \frac{1}{2i} \left( \hat G_\h - \hat G_\h^\dagger \right) \,.
\end{equation}

The elements of $\hat\rho_\h$ then form an infinite dimensional matrix whose positivity properties will be analyzed now. When taking into account the property of Eq.~(\ref{eq:Deltapm}), the matrix $\hat\rho_\h$ turns out to have a factorizable form, i.e. one finds the following explicit expressions for the spectral function
\begin{equation}
( \hat \rho_\h)_y^{\,y'}=\hat\rho_y\cdot \hat\rho_{\,y'}\,,   \label{eq:rhoab} 
\end{equation}
where, for $p^2\ge m_g^2$,
\begin{equation}
\hspace{-0.2cm} \hat\rho_{y} = \sqrt{\frac{2}{3\pi \rho R (1+R^2)}} \frac{1}{(1 - \bar y)^{1/2}} \IImm \left( (1 + iR) \left(1 -\bar y \right)^{\frac{3}{2} i R} \right) \,, \label{eq:rhoz} 
\end{equation}
with 
\begin{equation}
R(p) = \sqrt{(4/9) \cdot p^2/\rho^2 -1} \,. \label{eq:R}
\end{equation}

Given the factorization property, Eq.~(\ref{eq:rhoab}), it turns out that the operator $\hat \rho_\h$ is positive semidefinite: all its eigenvalues are zero except one $\lambda(p)$ (i.e.~$\det\hat\rho_\h=0$), which is given by the trace of the matrix, i.e.
\be
\lambda(p)=\tr\hat\rho_\h=\frac{1}{\rho}\int_0^{1}\bar \rho_\h(\bar y,\bar y;s) d\bar y \,.
\label{eq:lambda}
\ee
Using Eq.~(\ref{eq:rhoab}) one can see that there is a divergence at the value $\bar y=1$ so that the expression for $\lambda$ needs to be regularized. We will do it by introducing the cutoff $\bar\epsilon$ in the integral (\ref{eq:lambda}) which now will extend from $0$ to $1-\bar\epsilon$, so that the integral will be dominated by its value at $1-\bar\epsilon$, giving a term $\propto (-\log\bar\epsilon)$ which will be the leading one. As we will see this divergence will cancel out when computing physical observables, so it will not require any renormalization procedure. It turns out that $\lambda(p)$ is computed as
\be
\lambda(p)=\delta(p^2)+\left[ - \frac{\log\bar\epsilon}{2\pi\rho} \lambda_{\rm un}(s)+\mathcal O(\bar\epsilon^0)
   \right],\quad \lambda_{\textrm{un}}(s)=(s-m_g^2)^{-1/2} \,,
\ee
where $\delta(p^2)$ is the contribution to the spectral function of the graviton zero mode~\footnote{Its correct normalization comes from the prefactor in Eq.~(\ref{eq:rho_delta}) as $3\rho\int_0^{y_s}e^{-2A(y)}=1 $.} and $\lambda_{\rm un}(s)$ the contribution to the spectral function from the continuum, or unparticle contribution with a mass gap $m_g$ and dimension $d_U=3/2$~\cite{Delgado:2008gj}.

Therefore, in the diagonal basis the matrix $\hat \rho_\h^{\rm\, d}$ has all elements null except one, which can be chosen to be the element with $y=y'=y_s$, which is equal to $\lambda$, i.e.
\be
\left( \hat\rho_\h^{\rm\, d} \right)_y^{y'}=\lambda(p)\delta_{y, y_s}\delta_{y_s,y'} \,,
\ee
and the (infinite) orthogonal rotation $O_y^{y'}$ from $\hat\rho_\h\to \hat\rho_h^{\rm\, d}$ has, in particular, elements
\be
O^{y_s}_y=\frac{\hat\rho_y}{\sqrt{\lambda}}  \,.
\ee

Now the contribution of $\hat\rho_\h$ to a physical process where $\phi(y)$ is the profile along the extra dimension of the initial state in the tensor $T_{\mu\nu}(x,y)$, with a coupling for fields $\bar h_{\mu\nu}(x,y)$ given by
\be
\mathcal L_{5D}=-\frac{1}{\sqrt{2}M_5^{3/2}} e^{A(y)} \bar h^{\mu\nu}(x,y)T_{\mu\nu}(x,y) \,,
\ee
and $\psi(y')$ the profile of the final state in $T_{\rho\sigma}(x,y')$, is given by the element
\begin{align}
&\tr(\phi^T e^A\cdot \hat\rho_\h\cdot e^A \psi)=\tr(\phi^T e^A O^T\cdot \hat\rho_\h^{\rm\, d} \cdot O e^A\psi)\nonumber\\
&
=\sum_{y,y'}\phi_y e^{A_y}\left( O^T\right)^y_{y_s}\left(\hat\rho_\h^{\rm\, d}\right)_{y_s}^{y_s}O^{y_s}_{y'} e^{A_{y'}}\psi^{y'}
 =\sum_{y,y'}\frac{\phi_y  e^{A_y}\hat\rho_y}{\sqrt{\lambda}}\cdot \lambda \cdot \frac{\hat\rho_{y'}e^{A_{y'}}\psi^{y'}}{\sqrt{\lambda}}\nonumber\\
&=\int_0^{y_s} dy dy' \,\phi(y)e^{A(y)} \bar\rho_\h(y,y')e^{A(y')}\psi(y')=\int_0^{y_s} dy dy' \,\phi(y) \rho_\h(y,y')\psi(y')  \,,
\label{eq:expansion}
\end{align}
where we see that the divergence in the calculation of the eigenvalue $\lambda$ cancels out, while the last equality is written in terms of the 
spectral density $\rho_\h$ in the basis $h_{\mu\nu}(x,y)$. In particular if the initial state function is located at the brane $y=y_\alpha$, and the final state is localized at the brane $y=y_\beta$, then $\phi(y)\propto \delta(y-y_\alpha)$,
$\psi(y')\propto \delta(y'-y_\beta)$, and the result of Eq.~(\ref{eq:expansion}) is given by
\be
\tr(\phi^Te^A\cdot \hat\rho_\h\cdot e^A\psi)=e^{A(y_\alpha)+A(y_\beta)}\bar\rho_\h(y_\alpha,y_\beta;p)=\rho_\h(y_\alpha,y_\beta) \,.
\ee
The functions $\bar\rho_\h(y_\alpha,y_\beta;p)$ are those plotted in Fig.~\ref{fig:spectral_graviton}.

\section{Coupling of the graviton with SM matter fields}
\label{subsec:coupling_to_matter}

We are assuming fields located in the brane $y=y_\alpha$. Then, the usual form of the interaction Lagrangian in the 4D effective theory is given by the Lagrangian
\begin{equation}
\mathcal L_{5D} = -\frac{1}{\sqrt{2}M_5^{3/2}} T^{\mu\nu}(x,y) h_{\mu\nu}(x,y) \delta(y-y_\alpha)\,,
\label{eq:coupling}
\end{equation}
where $T_{\mu\nu}(x,y_\alpha)$ is the energy-momentum tensor of the matter fields localized at $y_\alpha$~\footnote{We are assuming here the simplified case where matter lives in some brane, as e.g.~the SM which is living in the IR brane, or perhaps some dark sector which could live in the UV brane. For matter (SM singlets) propagating in the extra dimension one should replace the interaction term in Eq.~(\ref{eq:coupling}) by $\int dy \, T^{\mu\nu}(x,y)h_{\mu\nu}(x,y)$.}. 

In particular, for the SM fields living in the IR brane at $y=y_1$, the energy-momentum tensor is given by
\be
T_{\mu\nu}=D_\mu H^\dagger D_\nu H+i\bar\psi \gamma_\mu D_\nu\psi - F_\mu{}^\rho F_{\nu\rho}-\eta_{\mu\nu}\mathcal L_{\rm SM} \,,
\label{eq:Tmunu}
\ee
where $D_\mu$ is the SM covariant derivative, $\psi$ corresponds to all SM left and right-handed fermions and $F_{\mu\nu}$ is the field strength of different gauge fields $F_{\mu\nu}=W_{\mu\nu}^a,B_{\mu\nu}$. The term proportional to $\eta_{\mu\nu}$ does not contribute to the different vertices as the tensor $h_{\mu\nu}$ is traceless. Notice that in the broken phase, when $\langle H\rangle=v/\sqrt{2} (0,1)^T$ the massive gauge bosons have contributions to $T_{\mu\nu}$ proportional to $m_V^2 V_\mu V_\nu$.
The graviton zero mode wave function, $h^0_{\mu\nu}(x,y)= h_0(y)h^0_{\mu\nu}$, where $ h_0(y)=\sqrt{3\rho}$ is canonically normalized as 
$\int_0^{y_s}dy\,e^{-2A}h_0^2 =1$, couples with the energy-momentum tensor at the IR brane as
\be
-\frac{1}{M_{\rm Pl}} T^{\mu\nu}(x,y_1)h_{\mu\nu}^0(x)\,.
\ee

We will now consider the coupling with matter of the continuum of KK modes, with Green's function $\mathcal{G}_\h(y,y')$. The effective field theory (EFT) for matter localized at the brane $y_\alpha$, for momenta $p\ll \rho$, provides the dimension eight operator $\mathcal O_\h(x,y_\alpha)$ with Wilson coefficient $c(y_\alpha)$ as
\begin{align}
\mathcal L_{\rm EFT}(y_\alpha)&=c(y_\alpha) \mathcal O_\h(x,y_\alpha),\nonumber\\
 \mathcal O_\h(x,y_\alpha)&=T^{\mu\nu}(x,y_\alpha)D_{\mu\nu,\rho\sigma}T^{\rho\sigma}(x,y_\alpha)=T^{\mu}_{\ \nu} T_{\ \mu}^{\nu} -\frac{1}{3} (T^{\mu}_{\ \mu})^2 \,,
 \label{eq:LEFT}
\end{align}
where the Wilson coefficients here have mass dimension $-4$. 

In particular, for the SM which is localized in the IR brane
\be
c(y_1)=-\frac{1}{6} \frac{1}{\rho^4} \,.
\ee
Thus gravitational interactions are suppressed by the TeV scale $\rho$, reflecting the fact that the continuum of KK modes is localized toward the IR. On the contrary, if there is some extra matter localized in the UV brane, SM singlets, the corresponding Wilson coefficient would be
\be
c(y_0)=-\frac{1}{9}\frac{1}{\rho^2 M_{\rm Pl}^2}\,.
\ee
In fact if we define the effective coupling $g_{\rm eff}(y_\alpha)$ as 
\be
|c(y_\alpha)|\equiv g_{\rm eff}^2(y_\alpha)\frac{1}{\rho^2} \,,
\ee
we can see that $g_{\rm eff}(y_0)\simeq 1/M_{\rm Pl}$ while $g_{\rm eff}(y_1)\simeq 1/\rho$.

\subsection{Low energy constraints}
The effective Lagrangian in Eq.~(\ref{eq:LEFT}) does give rise, in particular, to the dimension eight operators, in the notation of Refs.~\cite{Eboli:2006wa,Almeida:2020ylr},
\be
\mathcal L_{\rm EFT}\supset \sum_{i=0}^2\frac{f_{S_i}}{\rho^4}\mathcal O_{S_i} \,,
\ee
where
\begin{align}
\mathcal O_{S_0}&=(D^\mu H^\dagger D^\nu H)(D_\mu H^\dagger D_\nu H),\quad \mathcal O_{S_1}=(D^\mu H^\dagger D_\mu H)(D^\nu H^\dagger D_\nu H) \,, \nonumber \\
\mathcal O_{S_2}&=(D^\mu H^\dagger D^\nu H)(D_\nu H^\dagger D_\mu H)\,, \nonumber\\
f_{S_0}&=-\frac{1}{12},\quad f_{S_1}=\frac{1}{18}\,,\quad f_{S_2}=-\frac{1}{12}\,.
\label{eq:operatorsS}
\end{align}
The contributions of these effective operators to the observables $S,T,U$ have been computed in Ref.~\cite{Eboli:2006wa} as~\footnote{We thank Prof.~O.~J.~P.~\' Eboli for a private communication on this result.}
\be
\alpha T=-\frac{15}{16\pi^2}(m_W/\rho)^4\left(f_{S_0}+f_{S_2}+\frac{2}{5}f_{S_1}\right)(1+c_W^2)\frac{s_W^2}{c_W^2}\log(\rho/m_W) \,,
\ee
where $\alpha$ is the fine structure constant, and $s_W (c_W)$ the sine (cosine) of the electroweak mixing angle $\theta_W$, while $S=U=0$. Using now the values in Eq.~(\ref{eq:operatorsS}) we get
\be
\alpha T\simeq \frac{13}{96 \pi^2}\left(m_W/\rho \right)^4(1+c^2_W)\frac{s^2_W}{c^2_W}\log (\rho/m_W) \,,
\ee
which provides a very mild bound on the value of $\rho$ as $\alpha T\lesssim 10^{-5}$ ($10^{-6}$) for $\rho\gtrsim 500$ GeV (1 TeV). The small value of the $T$ parameter comes mainly because this effect stems from a dimension eight operator, and thus is suppressed by the fourth power of $1/\rho$.

\subsection{High energy constraints}

The effective Lagrangian in Eq.~(\ref{eq:LEFT}) does also give rise to a number of dimension eight operators, which contribute to an anomalous quartic gauge coupling (aQGC) as
\be
\mathcal L_{\rm EFT}\supset \sum_j \frac{f_{T_j}}{\rho^4}\mathcal O_{T_j}+\sum_k \frac{f_{M_k}}{\rho^4}\mathcal O_{M_k} \,,
\ee
where, using the notation of Refs.~\cite{Eboli:2006wa,Almeida:2020ylr}, we have
\begin{align}
\mathcal O_{T_0}&=(W^{\mu\nu}W_{\mu\nu})(W_{\alpha\beta}W^{\alpha\beta}),\ \mathcal O_{T_2}=(W^{\mu\alpha}W_{\nu\alpha})(W_{\mu\beta}W^{\nu\beta}) \,, \nonumber\\
\mathcal O_{T_5}&=(W^{\mu\nu}W_{\mu\nu})(B_{\alpha\beta}B^{\alpha\beta}),\ \hspace{0.17cm} \mathcal O_{T_7}=(W^{\mu\alpha}W_{\nu\alpha})(B_{\mu\beta}B^{\nu\beta}) \,, \nonumber\\
\mathcal O_{T_8}&=(B^{\mu\nu}B_{\mu\nu})(B_{\alpha\beta}B^{\alpha\beta}),\ \hspace{0.34cm} \mathcal O_{T_9}=(B^{\mu\alpha}B_{\nu\alpha})(B_{\mu\beta}B^{\nu\beta}) \,,
\label{eq:operatorsT}
\end{align}
and 
\begin{align}
&\mathcal O_{M_0}=(W^{\mu\nu}W_{\mu\nu})(D^\alpha H^\dagger D_\alpha H),\ \mathcal O_{M_1}=(W^{\mu\alpha}W_{\nu\alpha})(D_\mu H^\dagger D^\nu H) \,, \nonumber\\
&\mathcal O_{M_2}=(B^{\mu\nu}B_{\mu\nu})(D^\alpha H^\dagger D_\alpha H),\ \hspace{0.18cm}\mathcal O_{M_3}=(B^{\mu\alpha}B_{\nu\alpha})(D_\mu H^\dagger D^\nu H) \,,
\label{eq:operatorsM}
\end{align}
with Wilson coefficients
\begin{align}
&f_{T_0}=\frac{1}{18},\, f_{T_2}=-\frac{1}{6},\, f_{T_5}=\frac{1}{9},\,
f_{T_7}=-\frac{1}{3},\, f_{T_8}=\frac{1}{18},\, f_{T_9}=-\frac{1}{6} \,, \nonumber\\
&f_{M_0}=-\frac{1}{9},\, f_{M_1}=\frac{1}{3},\, f_{M_2}=-\frac{1}{9},\,
f_{M_3}=\frac{1}{3} \,.
\end{align}

The LHC constraints on the above operators are obtained from the CMS experiment~\cite{Sirunyan:2019der,Sirunyan:2020tlu,Sirunyan:2020gyx}. The strongest constraints are over the operators $\mathcal O_{T_0}$ and $\mathcal O_{T_2}$ which translate into the 95\% CL lower bound $m_g\gtrsim 1.3$ TeV. Projections in FCC-hh, at $\sqrt{s}=100$ TeV and integrated luminosities up to 30 ab$^{-1}$, have been made on the anomalous $WW\gamma\gamma$ couplings~\cite{Ari:2021ixv} which, for leptonic decay channels of the $W$s in the final state, yield future bounds reaching values as $m_g\gtrsim 7$ TeV.

Of course the presence of aQGC induces violation of unitarity, e.g.~in longitudinal gauge boson scattering processes involving four vector particles, as the corresponding scattering amplitudes grow with $\hat s^2$, where $\sqrt{\hat s}$ is the center-of-mass energy, since the SM cancellation fails. This issue has been generally considered for the operators $\mathcal O_{S_i}$, $\mathcal O_{T_j}$ and $\mathcal O_{M_k}$ in Refs.~\cite{Almeida:2020ylr,Guo:2020lim}. The unitarity violation indicates a failure of the EFT to describe the corresponding processes at such large values of $\sqrt{\hat s}$. In particular using the general results in Ref.~\cite{Almeida:2020ylr} the unitarity constraints imply an upper bound as $\sqrt{\hat s}\lesssim 2\, m_g$ for the validity of the EFT.

\vspace{1cm}
\section{The radion}
\label{sec:radion}

The radion field $F(x,y)$ is defined as the scalar perturbation of the metric
\begin{align}
ds^2 &= e^{-2A(y) - 2F(x,y)} \eta_{\mu\nu} dx^\mu dx^\nu - [1 + G(x,y)]^2 dy^2 \,, \label{eq:metric_radion}\nonumber \\
\phi(x,y) &= \phi(y) + \psi(x,y) \,,
\end{align}
with $F(x,y) = F(y) \mathcal R(x)$. When considering an appropriate gauge choice, the EoM for the $y$-dependent part become~\cite{Csaki:2000zn}
\begin{align}
&\left(p^2 e^{2A} A^{\prime\prime}(y)^{-1} - 2 \right) F(y) + \frac{d}{dy}\left[  e^{2A} A^{\prime\prime}(y)^{-1} \partial_y \left( e^{-2A} F(y) \right)  \right]  = 0 \,, \nonumber \\
&F^\prime - 2 A^\prime F = \phi^\prime \psi  \,, \quad
G = 2F \,. \label{eq:Gradion}
\end{align}
After rescaling the field by $F(z) = e^{3A(z)/2} \phi^\prime(z) \tilde F(z)$, one can cast the EoM in a Schr\"odinger-like form, as
\be
- { \tilde F}^{\prime\prime}(z) + V_F(z) \tilde F(z) = p^2 \tilde F(z) \,,
\ee
where the potential is given by
\begin{equation}
V_F(z) = \frac{9}{4} {A^\prime}^2(z) + \frac{5}{2} A^{\prime\prime}(z) - A^\prime(z) \frac{\phi^{\prime\prime}(z)}{\phi^\prime(z)} - \frac{\phi^{\prime\prime\prime}(z)}{\phi^\prime(z)} + 2 \left( \frac{\phi^{\prime\prime}(z)}{\phi^\prime(z)} \right)^2 \,.
\end{equation}
This potential turns out to be equal to the constant value $V_F(z) = m_g^2$, where $m_g = 3\rho/2$ is the mass gap for the radion. This value of the mass gap equals that of the graviton in previous sections.

\subsection{The radion Green's functions}
\label{subsec:GF_radion}

After making the field redefinition $F(x,y)\to \kappa F(x,y)$, as for the case of the graviton, the EoM for the radion Green's function $G_F(y,y^\prime;p)$~\cite{Csaki:2000zn,Megias:2015ory} is the same as the one for the graviton, cf. Eq.~(\ref{eq:eom_graviton3}). After fixing the value of $y^\prime$, we can divide the $y$ space into the following domains: $0 \le y \le y^\prime$ and $y^\prime \le y \le y_s$. Then, the general solution is
\begin{equation}
G_F(y,y^\prime;p) =  \left\{ 
\begin{array}{cc}
C^{I}_1 \cdot (y_s - y)^{\frac{3}{2}\Delta_F^-} + C^{I}_2 \cdot (y_s - y)^{\frac{3}{2} \Delta_F^+} & \quad y  <   y^\prime < y_s  \\
C^{II}_1 \cdot (y_s - y)^{\frac{3}{2}\Delta_F^-} + C^{II}_2 \cdot (y_s - y)^{\frac{3}{2} \Delta_F^+} & \quad  y^\prime < y < y_s  
\end{array} \,, \right. \label{eq:GF_general}
\end{equation}
where we have defined $\Delta_F^\pm = \Delta_\h^\pm$ and $\delta_F = \delta_\h$, cf. Eq.~(\ref{eq:Delta_h}). The Green's function is subject to boundary and matching conditions in the UV and IR branes, as well as for $y = y^\prime$. These read
\begin{align}
\begin{split}
& G_F^\prime(0,y^\prime) =   \left( \frac{1}{3}\kappa^2 W(\phi(y)) - \frac{2 p^2 e^{2A(y)}}{U_0^{\prime\prime}(\phi(y))}  \right)  G_F(y) \Bigg|_{y = 0}  \,, \\ 
& \Delta G_F(y^\prime,y^\prime) = 0 \,, \qquad \Delta G_F^\prime(y^\prime,y^\prime) = e^{4A(y^\prime)}  \,, \\
& \Delta G_F(y_1,y^\prime) = 0 \,, \qquad \Delta G_F^{\prime}(y_1,y^\prime) =  0   \,,  \label{eq:GF_bc}
\end{split}
\end{align}
where the localized effective potential in the UV brane $U_0(\phi)$ is defined by Eq.~(\ref{eq:U0}), and its second derivative turns out to be $U_0^{\prime\prime}(\phi(0)) = \gamma_0 - 2\rho$, which in the following we will denote by $U_0^{\prime\prime}$. In addition, we should impose regularity in the IR, i.e.~we consider $C_1^{II} = 0$. After implementing the boundary and matching conditions in the general solution, one finds
\begin{eqnarray}
G_F(y,y^\prime) &=& \frac{1}{3\rho} \frac{1}{\delta_F} (1- \bar y_\uparrow)^{\frac{3}{2}\Delta_F^+} \label{eq:GF} \\
&&\qquad \times \left[ - (1 - \bar y_\downarrow)^{\frac{3}{2}\Delta_F^-}  +   \left( 1 +\frac{3 U_0^{\prime\prime}}{2\rho} \frac{\delta_F}{\Phi(p)} \right)  (1-\bar y_\downarrow)^{\frac{3}{2}\Delta_F^+} \right] \,, \nonumber
\end{eqnarray}
where
\begin{align}
\Phi(p)&= \frac{p^2}{\rho^2} - \frac{U_0^{\prime\prime}}{4\rho} (1 + 3 \delta_F) \,. \label{eq:Phi}
\end{align}
The analytical expressions of the brane-to-brane Green's functions are 
\begin{eqnarray}
G_F(y_0,y_0;p) &=&  \frac{U_0^{\prime\prime}}{2\rho^2} \frac{1}{\Phi(p)} \,, \label{eq:GF_y0y0}  \\
G_F(y_0,y_1;p) &=&  \frac{U_0^{\prime\prime}}{2\rho^2} \frac{1}{\Phi(p)} e^{-\frac{3 A_1}{2}\Delta_F^+} \,,  \label{eq:GF_y0y1}  \\
G_F(y_1,y_1;p) &=&  \frac{1}{3\rho} \frac{1}{\delta_F} e^{3A_1}  \left[ -1 + \left( 1 + \frac{3 U_0^{\prime\prime}}{2\rho} \frac{\delta_F}{\Phi(p)} \right) e^{-3 A_1\delta_F} \right] \,, \label{eq:GF_y1y1} 
\end{eqnarray}
and their low momentum behaviors are
\begin{eqnarray}
G_F^{-1}(y_0,y_{0,1})  &\stackrel[ p \ll \rho ]{\simeq}{}&  -2\rho   + \mathcal O(p^2)  \,,  \label{eq:GF_y0y01_asymp} \\
G_F^{-1}(y_1,y_1) &\stackrel[ p \ll \rho ]{\simeq}{}&  -\frac{6\rho^4}{2 k^3 + \rho^3} + \mathcal O(p^2)  \,. \label{eq:GF_y1y1_asymp} 
\end{eqnarray}
In the following we will denote the zero momentum limits of the brane-to-brane Green's functions as $G_F^{\alpha\beta} \equiv  \lim_{p \to 0} G_F(y_\alpha,y_\beta;p)$.

\begin{figure}[t]
\centering
\includegraphics[width=4.6cm]{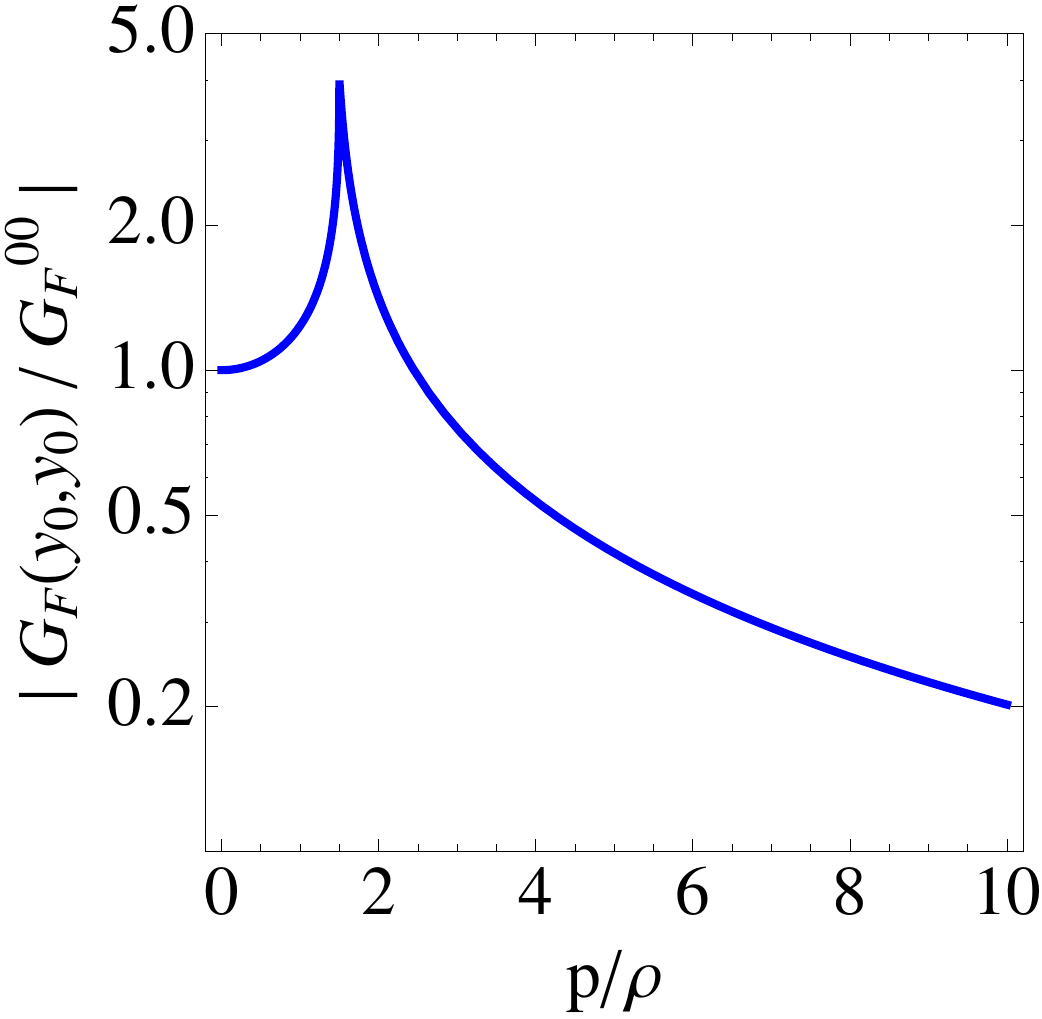}  \hspace{0.3cm}
\includegraphics[width=4.35cm]{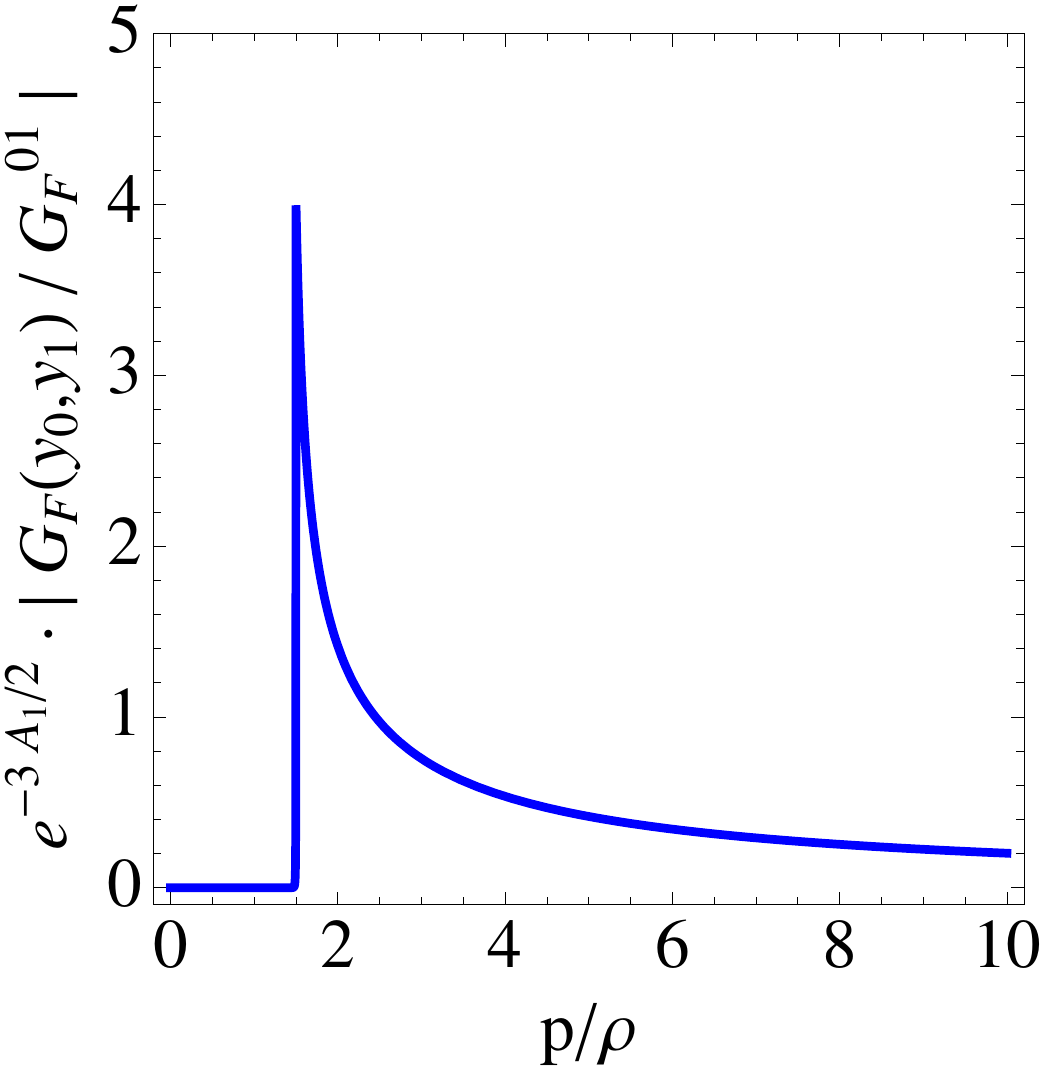}  \hspace{0.3cm}
\includegraphics[width=4.9cm]{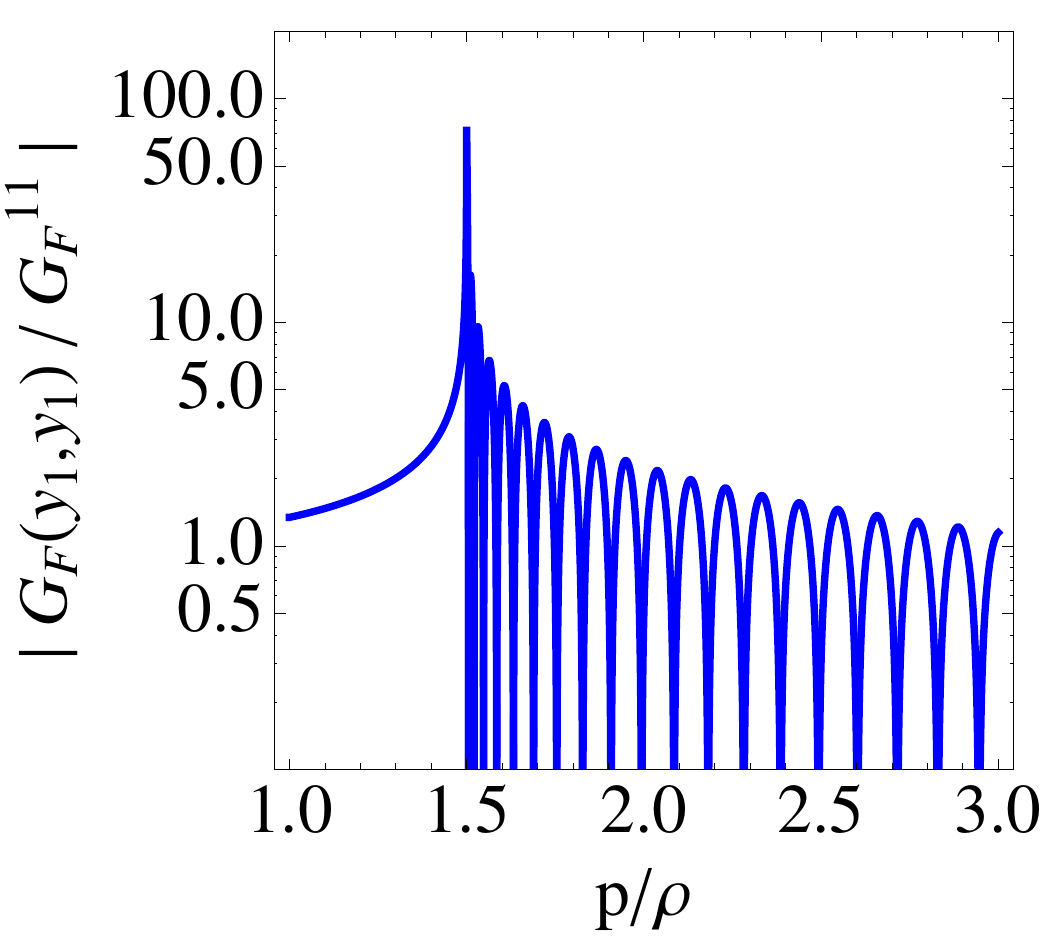}
\caption{\it Plots of $|G_F(y_0,y_0;p)/G_F^{00}|$ (left panel), $|G_F(y_0,y_1;p)/G_F^{01}|$ (middle panel), 
and $|G_F(y_1,y_1;p)/G_F^{11}|$ (right panel) as functions of $p/\rho$. We have used $A_1 = 23$ and $U_0^{\prime\prime} = k$ in all panels, and assume time-like momenta $p^2>0$.
}
\label{fig:Gradion}
\end{figure}

\begin{figure}[t]
\centering
\includegraphics[width=4.6cm]{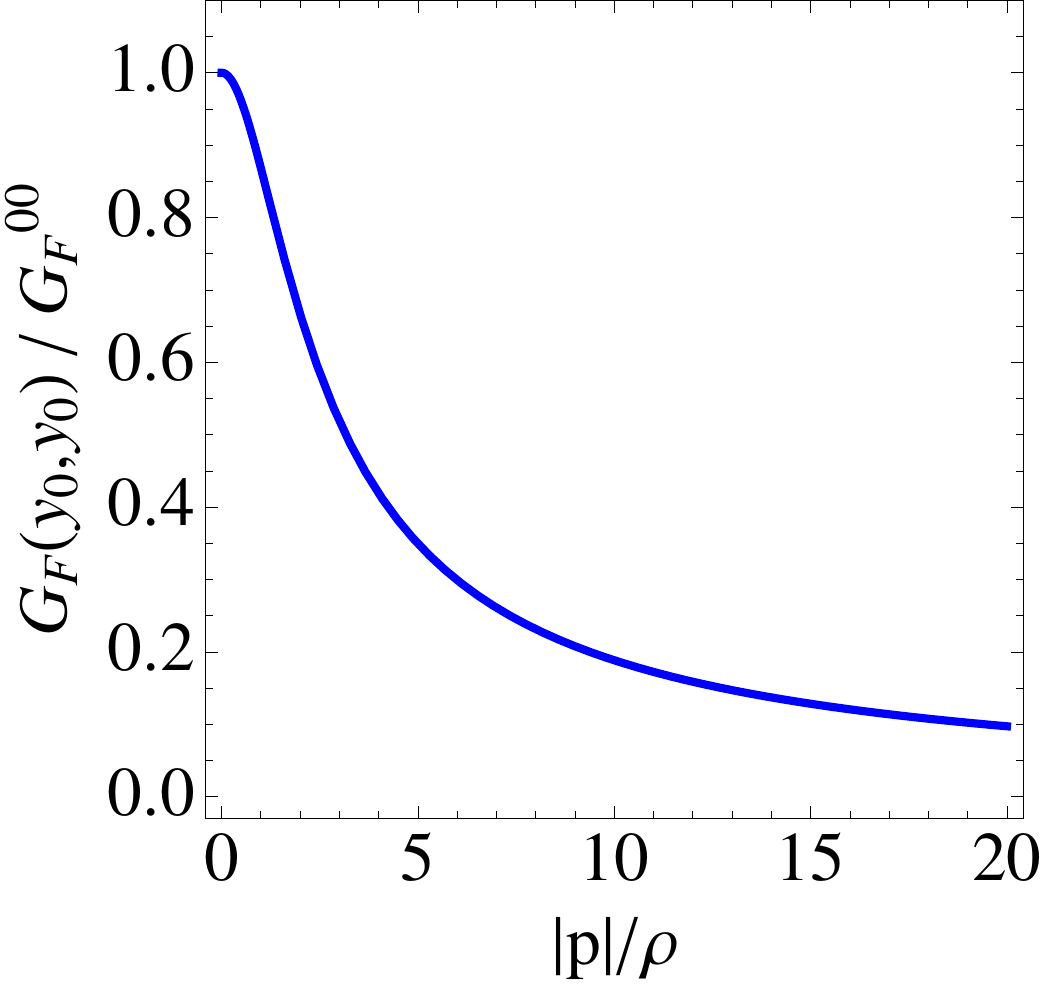}  \hspace{0.3cm}
\includegraphics[width=4.5cm]{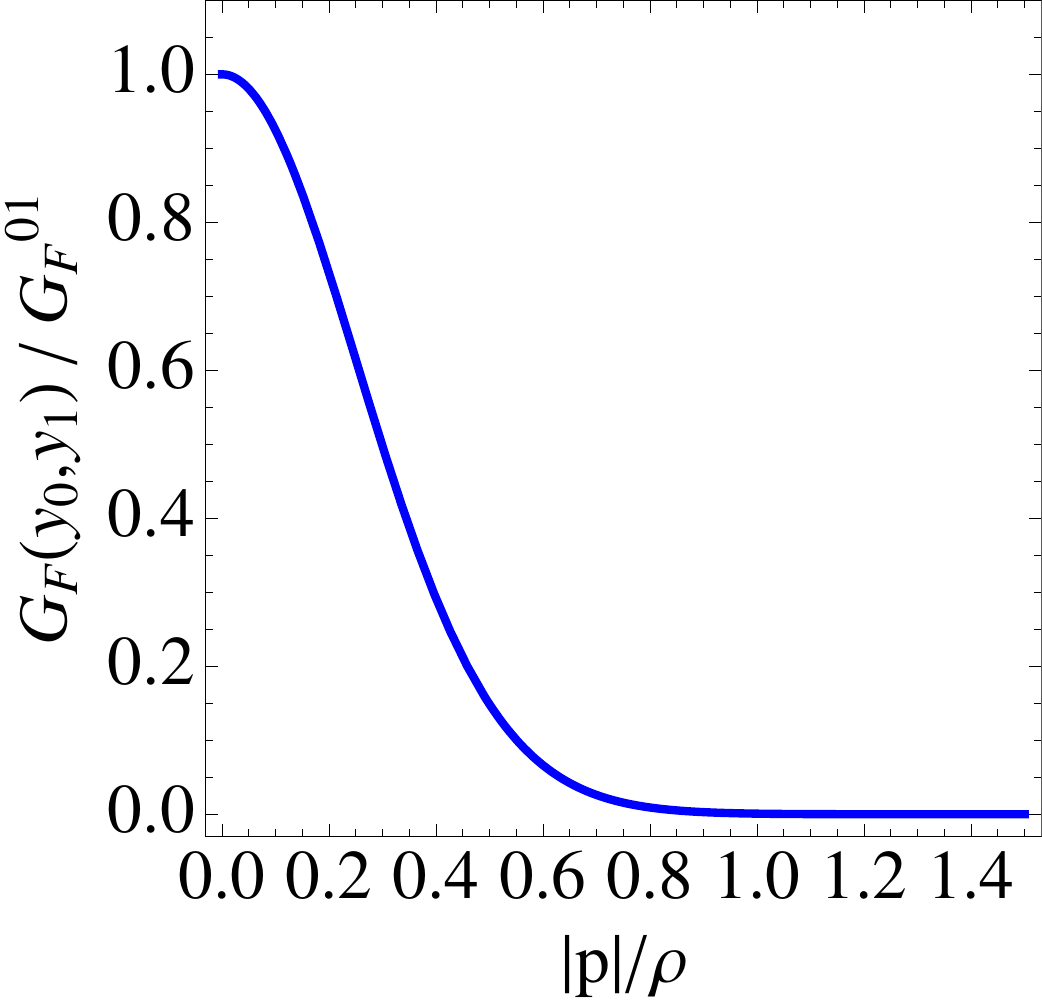}  \hspace{0.3cm}
\includegraphics[width=4.6cm]{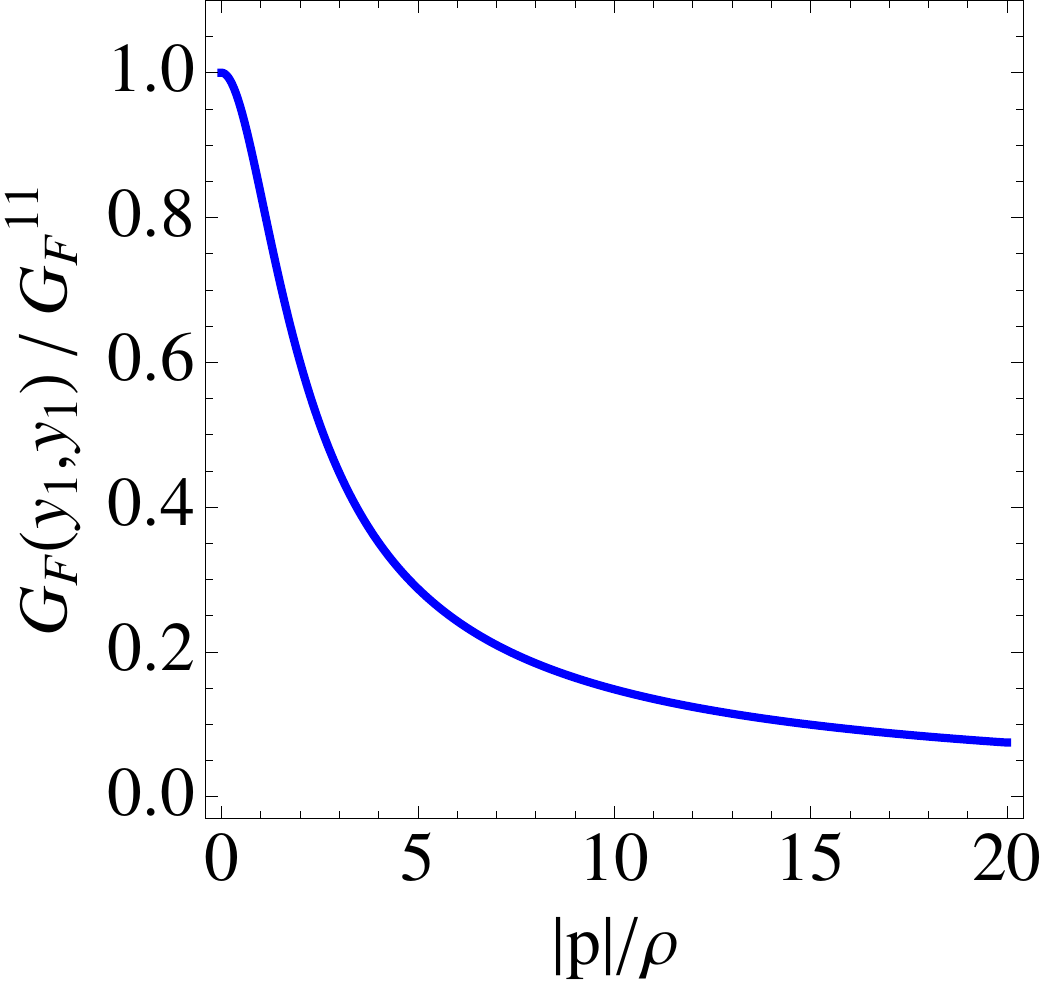}
\caption{\it Plots of $G_F(y_0,y_0;|p|)/G_F^{00}$ (left panel), $G_F(y_0,y_1;|p|)/G_F^{01}$ (middle panel), and $G_F(y_1,y_1;|p|)/G_F^{11}$ (right panel) as functions of $|p|/\rho$. We have used $A_1 = 23$ and $U_0^{\prime\prime} = k$ in all panels, and assume space-like momenta $p^2<0$.
}
\label{fig:Gradion_pE}
\end{figure}

We plot in Fig.~\ref{fig:Gradion} the result for the Green's functions $G_F(y_0,y_0)$, $G_F(y_0,y_1)$  and $G_F(y_1,y_1)$, normalized to their zero momentum limits, as functions of $p/\rho$, for time-like momenta $p^2 > 0$. For space-like momenta $p^2 < 0$ the Green's functions are purely real. We plot in Fig.~\ref{fig:Gradion_pE} the Green's functions as functions of $|p|/\rho$, in the latter case.

\begin{figure}[t]
\centering
\includegraphics[width=5.3cm]{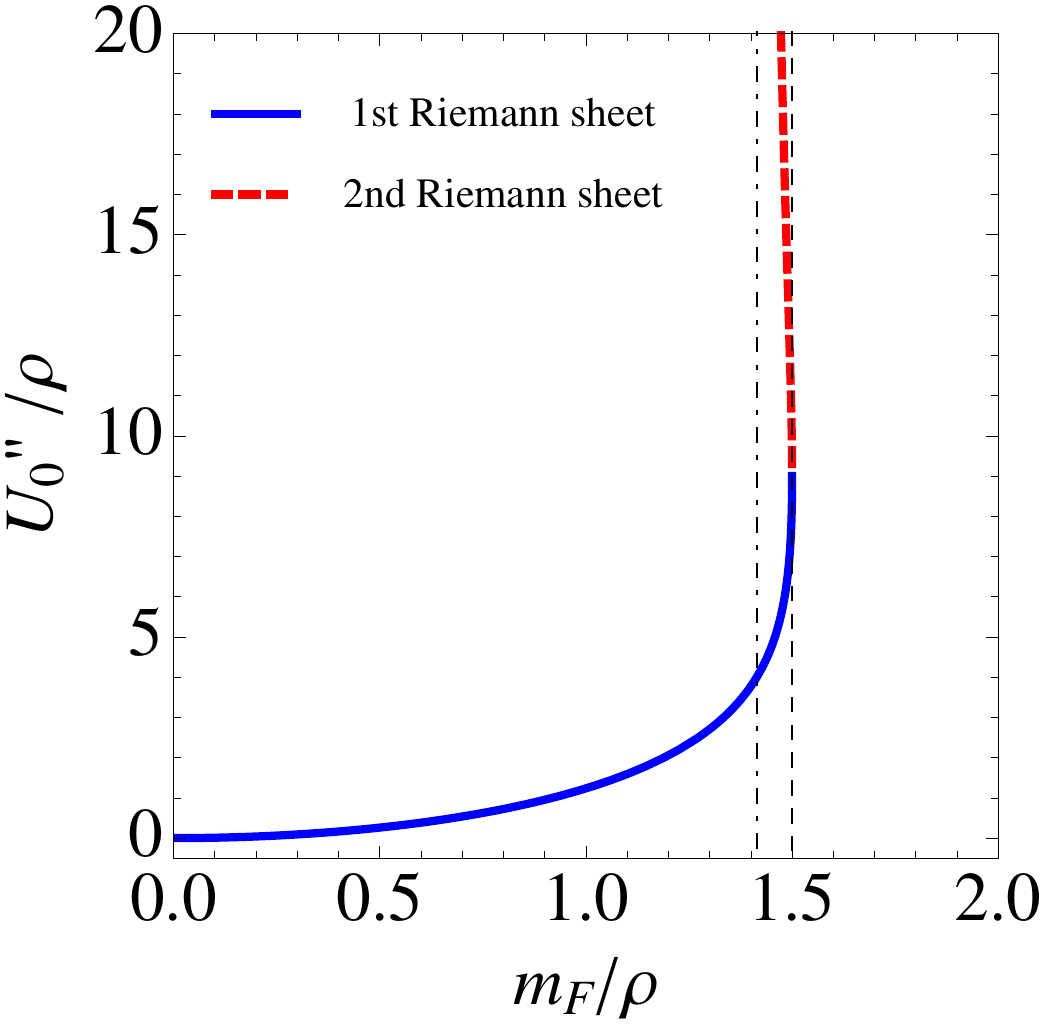} \hspace{1.5cm}
\includegraphics[width=4.9cm]{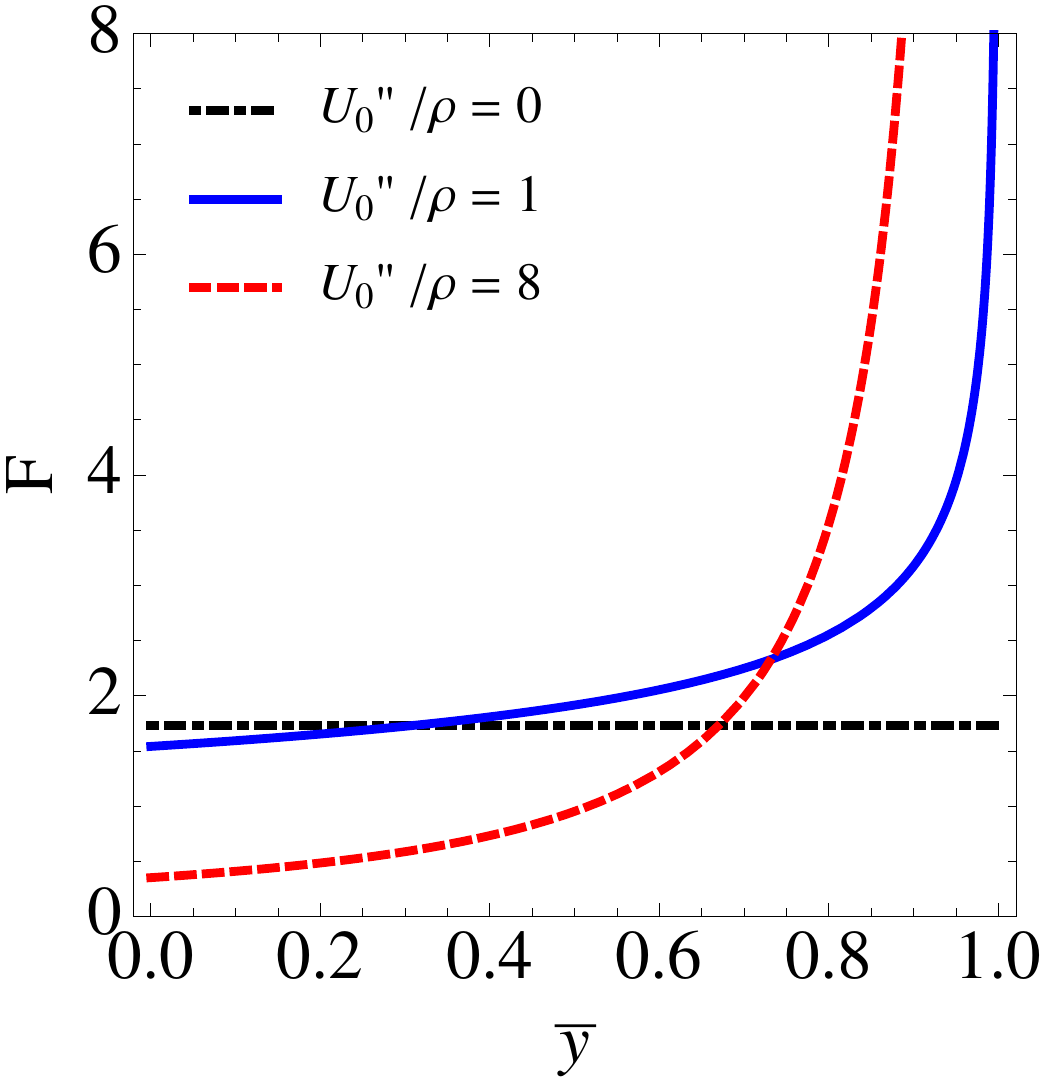}
\caption{\it Left panel: Parametric dependence of the isolated zero of $\Phi(m_F)$. We display $U_0^{\prime\prime}/\rho$ as a function of $m_F/\rho$ in the 1st Riemann sheet (solid blue) and in the 2nd Riemann sheet (dashed red) as given by Eq.~(\ref{eq:U0pp_zero}).  To guide the eye, we display vertical lines for values $m_F = \sqrt{2} \rho$ (dot-dashed black) and $m_F = m_g$ (dashed black). Notice that the range $0 \le U_0^{\prime\prime}/\rho \le 9$ corresponds to the pole of the Green's function in the 1st Riemann sheet, while the range $9 \le U_0^{\prime\prime}/\rho < \infty$ corresponds to the pole in the 2nd Riemann sheet. Right panel: Normalized wave function for the radion as a function of $\bar y := \rho y$, as given by Eq.~(\ref{eq:Fnorm}). We display the results for $U_0^{\prime\prime}/\rho = 0$, $1$ and $8$.
}
\label{fig:phi_zero_Fradion}
\end{figure}

Notice that (unlike the graviton case) the Green's functions do not have an isolated massless mode as their behavior in the limit $p\to 0$, as shown in Eqs.~(\ref{eq:GF_y0y01_asymp})-(\ref{eq:GF_y1y1_asymp}), yields a constant value and not an isolated singularity. This point is in agreement with previous studies in the subject in Ref.~\cite{Cabrer:2009we}. The function  $\Phi(p)$ given by Eq.~(\ref{eq:Phi}) has a single zero, either in the first or second Riemann sheet. Let us write the equation $\Phi(m_F)=0$, with $m_F$ the mass of the radion, in the form
\begin{equation}
\frac{U_0^{\prime\prime}}{\rho} = \frac{4 m_F^2/\rho^2}{1 + 3\delta_F} \quad \textrm{with} \quad \delta_F = \pm \sqrt{1-(4/9) \cdot m_F^2/\rho^2} \,,  \label{eq:U0pp_zero}
\end{equation}
where the $+ (-)$ corresponds to the first(second) Riemann sheet. We display in the left panel of Fig.~\ref{fig:phi_zero_Fradion} the parametric dependence of $U_0^{\prime\prime}/\rho$ with $m_F$ as given by Eq.~(\ref{eq:U0pp_zero}). One can see that $U_0^{\prime\prime} \ge 0$ demands that $m_F \le m_g$ when considering the 1st Riemann sheet, while $\sqrt{2} \rho < m_F \le m_g$ in the 2nd Riemann sheet, so that the mass is below the mass gap, except for $U_0^{\prime\prime}/\rho = 9$ where it has the same value. Notice that there exists a zero of $\Phi(m_F)$ in the first Riemann sheet only if $0 \le U_0^{\prime\prime}/\rho \le 9$, while the existence of a zero in the second Riemann sheet demands that $9 \le U_0^{\prime\prime}/\rho <  +\infty$. The zero of $\Phi(m_F)$, either in the first or second Riemann sheet, is given by
\begin{equation}
\frac{m_F^2}{\rho^2} = \frac{1}{8} \frac{U_0^{\prime\prime}}{\rho} \left[ -\left( \frac{U_0^{\prime\prime}}{\rho} - 2 \right) + \sqrt{ \left( \frac{U_0^{\prime\prime}}{\rho} - 2 \right)^2 + 32  } \right] \,. \label{eq:phi_zero}
\end{equation}
This zero corresponds to an isolated pole of the Green's function. Note that $m_F^2/\rho^2$ is always real and positive, hence this pole corresponds to a bound state of the radion spectrum that should be located in the first Riemann sheet, i.e. the physical sheet. Given the considerations above, we conclude that such bound state for the radion exists only for values $0 \le U_0^{\prime\prime}/\rho \le 9$~\footnote{Notice that when considering negative values of $U_0^{\prime\prime}$ or taking the negative sign in front of the square root of Eq.~(\ref{eq:phi_zero}), this equation predicts the existence of an unphysical tachyonic mode $(m_F^2 < 0)$.  One can see from Eq.~(\ref{eq:U0pp_zero}) that such a mode is in the second Riemann sheet for $U_0^{\prime\prime}/\rho > 0$, and in the first Riemann sheet for $U_0^{\prime\prime}/\rho < 0$.}.

\begin{figure}[t]
\centering
\includegraphics[width=4.6cm]{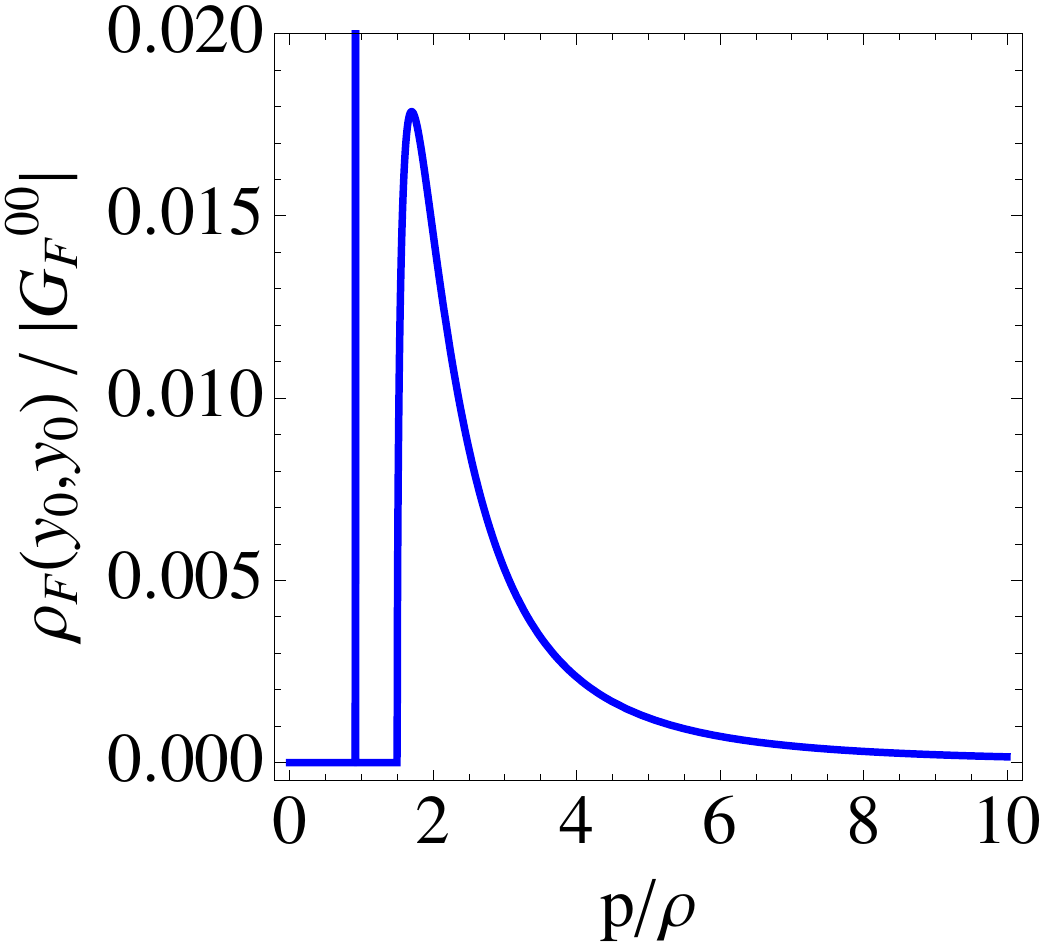}  \hspace{0.3cm}
\includegraphics[width=4.4cm]{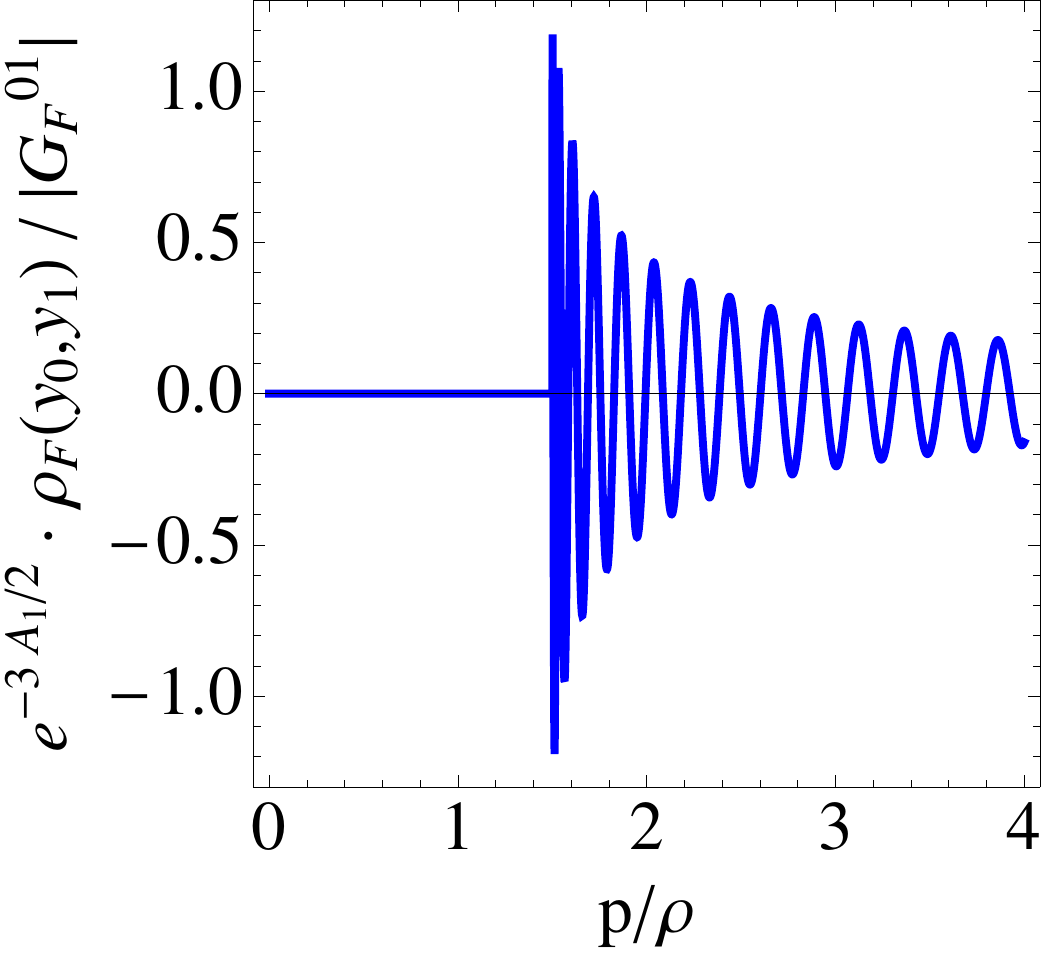} \hspace{0.3cm}
\includegraphics[width=4.1cm]{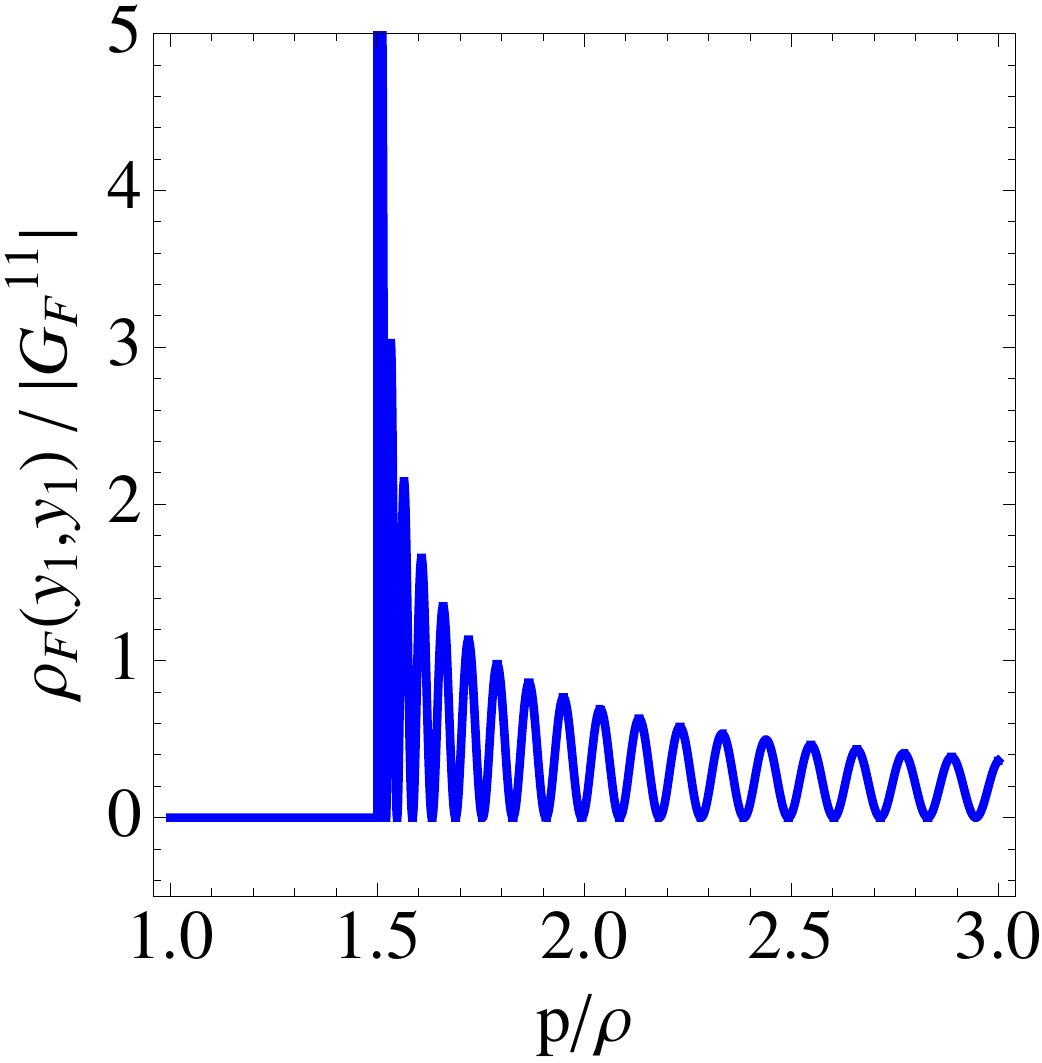} 
\caption{\it Spectral functions $\rho_F(y_0,y_0;p)$ (left panel), $\rho_F(y_0,y_1;p)$ (middle panel) and $\rho_F(y_1,y_1;p)$ (right panel) as a function of $p/\rho$, for a continuum radion. We have used $A_1 = 23$ and $U_0^{\prime\prime} = \rho$ (left panel) and $U_0^{\prime\prime} = k$ (middle and right panels), and assume time-like momenta $p^2>0$.
}
\label{fig:spectral_radion}
\end{figure}

Finally, let us study the spectral functions for the radion. In Fig.~\ref{fig:spectral_radion} we show $\rho_F(y_0,y_0;p)$, $\rho_F(y_0,y_1;p)$ and $\rho_F(y_1,y_1;p)$, as functions of $p/\rho$. Note that the isolated mode appears as a Dirac delta contribution. In addition, there appears a continuum for $p > m_g$. We are using the prescription $p^2 \to p^2 + i\epsilon$, so that for real values of $p$ above the mass gap, $p > m_g$, $\delta_F$ is computed in the physical sheet as $\delta_F = -i R$, where $R$ is given by Eq.~(\ref{eq:R}). Finally note that for the choice of $U^{\prime\prime}_0=k$, middle and right panels, no Dirac delta function appears in the corresponding spectral functions.

\subsection{The radion spectrum}
\label{sec:radion_spectrum}

In order to compute the spectrum of the radion, we have to solve the EoM of Eq.~(\ref{eq:Gradion}). The general solution of this equation is
\begin{equation}
F(y) =  C_{1F} \cdot (y_s - y)^{\frac{3}{2}\Delta_F^-} + C_{2F} \cdot (y_s - y)^{\frac{3}{2} \Delta_F^+}, \qquad  y  < y_s    \,. \label{eq:F}
\end{equation}
The wave function is subject to the following boundary condition in the UV brane and jumping conditions in the IR brane
\begin{eqnarray}
C_{\UV}(p)  &\equiv &  \frac{F^\prime(y)}{F(y)} \Bigg|_{y=0}  -  \left( \frac{1}{3}\kappa^2 W(\phi(y)) - \frac{2 p^2 e^{2A(y)}}{U_0^{\prime\prime}(\phi(y))}  \right) \Bigg|_{y=0} = 0 \,, \label{eq:F_bc_uv} \\
\Delta F(y_1)  &=& 0 \,, \qquad \Delta F^{\prime}(y_1) =  0 \,, \label{eq:F_matching_ir}
\end{eqnarray}
as well as regularity in the IR, which demands that $C_{1F} = 0$.  The integration constant $C_{2F}$ can be fixed by normalization of the wave function as we will see below. Notice that the wave function of Eq.~(\ref{eq:F}) fulfills by construction the IR brane conditions, but it remains the UV boundary condition which is fulfilled only for certain values of the momentum. The zeros of the function $C_{\UV}(p)$ will then lead to the spectrum of the radion. From an explicit computation of $C_{\UV}(p)$, the result for this function turns out to be
\begin{equation}
C_{\UV}(p) = 1/G_F(y_0,y_0;p)\,,
\end{equation}
where the explicit expression of the UV-to-UV Green's function is given by Eq.~(\ref{eq:GF_y0y0}). Then, we conclude that the radion spectrum contains a single bound state corresponding to the pole of the Green's function, and hence to the zero of the function $\Phi(m_F)$. The squared mass of this state is given by Eq.~(\ref{eq:phi_zero}), and exists only for values $0 \le U_0^{\prime\prime}/\rho \le 9$. Notice also that the function $C_{\UV}(p)$ in the regime $p \ll \rho$ behaves as
$C_{\UV}(p) = -2 \rho  + {\mathcal O}(p^2) \,, \label{eq:Cuv_approx_psmall}$
which is non-vanishing. This implies that the UV boundary condition is not fulfilled in this limit, and hence there is no massless mode for the radion. Finally, the wave function can be normalized to
\begin{equation}
\int_0^{y_s} dy \, e^{-2A(y)} |F(y)|^2 = y_s \,,
\end{equation}
leading to
\begin{equation}
F(y) = \sqrt{3 \delta_F}\,e^{\frac{3}{2}A(y)(1-\delta_F)} \,, \label{eq:Fnorm}
\end{equation}
with the value~$\delta_F = ( - U_0^{\prime\prime}/\rho + [ \left( U_0^{\prime\prime}/\rho - 2  \right)^2 + 32 ]^{1/2} )/6$. This result for the normalized wave function is displayed in the right panel of Fig.~\ref{fig:phi_zero_Fradion}. Notice that the radion is mostly localized toward the IR brane, as $F(y)$ is a monotonically increasing function and divergent in the limit $y \to y_s$. The localization is controlled by $U_0^{\prime\prime}$, so that the effect becomes more important for $U_0^{\prime\prime} /\rho \to 9$, while in the opposite limit, $U_0^{\prime\prime}/\rho \to 0$, the radion wave function becomes flat, as it should as the interval distance is not stabilized in that case.

\subsection{Coupling to the SM fields}
\label{subsec:coupling_radion_SM}
 
The coupling of the radion to the SM fields, localized on the IR brane, is with the trace of the energy-momentum tensor
\be
\mathcal L_{4D}=-\frac{1}{\sqrt{2} M_5^{3/2}}F(x,y_1)\mathcal T,\quad  \mathcal T\equiv\tr T^\mu_{\ \nu}(x,y_1)\,.
 \ee
 This gives rise, upon integration of the radion continuum, to the EFT Lagrangian with dimension eight operators as
 \be
 \mathcal L_{\rm EFT}=c_F(y_1) \mathcal O_F, \quad \mathcal O_F=\mathcal T^2 \,,
 \ee
 where the Wilson coefficient $c_F(y_1)$ is equal to
 \be
 c_F(y_1)\simeq -\frac{1}{6} \frac{1}{\rho^4} \,.
 \ee

The dimension eight operator $\mathcal T^2$ gives rise to a subset of the operators in Eqs.~(\ref{eq:operatorsS}), (\ref{eq:operatorsT}) and (\ref{eq:operatorsM}), in particular to $\mathcal O_{S_1}$, $\mathcal O_{M_0}$, $\mathcal O_{M_2}$, $\mathcal O_{T_0}$, $\mathcal O_{T_5}$ and $\mathcal O_{T_8}$, with corresponding Wilson coefficients $f_{S_1}=-1/6$, $f_{M_0}=1/3$, $f_{M_2}=1/3$, $f_{T_0}=-1/6$, $f_{T_5}=-1/3$, and $f_{T_8}=-1/6$.

Similarly to the case of the graviton continuum, the operator $\mathcal O_{S_1}$ gives rise to the observable $T$ as
\be
\alpha T=\frac{1}{16 \pi^2}\left(m_W/\rho \right)^4 (1+c_W^2)\frac{s_W^2}{c_W^2}\log(\rho/m_W) \,,
\ee
which is $\alpha T \lesssim 4\times 10^{-6}$  $(4\times 10^{-7})$ for $\rho\gtrsim 500$ GeV ($1$ TeV), and thus harmless for electroweak observables.

As for the high-energy constraints the analysis is similar to that performed for the case of the graviton continuum, including the violation of unitarity in gauge boson scattering processes. Similarly the isolated radion light mode also couples to the SM and its phenomenology depends to a large extent on its mass which, as we have seen is a function of the parameter $U^{\prime\prime}_0$. A detailed analysis of its phenomenology is beyond the scope of the present paper and will be done elsewhere.

\section{Conclusions}
\label{sec:conclusions}

Theories with a warped extra dimension, relating the 5D Planck scale and the electroweak scale by the warp factor, are among the best candidates to solve the hierarchy problem. In view of the strong constraints from the LHC direct searches of isolated narrow resonances, a possible solution to alleviate the experimental tension could be held in theories where the KK spectrum is a continuum, as no bump in cross sections should appear, but only indirect deviations from the Standard Model predictions.  

Linear dilaton models have recently received a lot of attention, as their UV completion is a type II string theory, where non-abelian gauge interactions arise non-perturbatively localized on stacks of NS5-branes, with gauge coupling geometrically determined by moduli of the 5-branes where they are confined, and thus independent of the value of the string coupling $g_s$. This allows to have string scales $M_s$ much smaller than the 4D Planck mass. This theory, in the decoupling limit $g_s\to 0$, is described by the Little String Theory, a string theory where gravity decouples. From the bottom-up approach dilaton models arise from a 5D model with a warped extra dimension and a critical  bulk potential, in terms of the stabilizing field, given by $V(\bar\phi)\propto e^{2\bar\phi}$, see Eq.~(\ref{eq:W_lineardilaton}). In proper coordinates $y$ the metric has a naked singularity at a finite value $y_s=1/|\rho|$, which corresponds in conformal coordinates $z$ to the limit $z_s\to \infty$. 

For $\rho<0$ one has to introduce an interval between two boundaries, at the UV ($y=0$) and at the IR ($y=y_1<y_s$), with a $\mathbb Z_2$ orbifold symmetry, and the theory has a discrete spectrum. In this case the string scale is of order the TeV scale, and the length of the interval is dictated by the 4D Planck scale, so that the SM fields are in the UV brane, the hierarchy problem is entirely solved by the string theory and the string coupling is tiny, $g_s\sim 10^{-15}$. For $\rho>0$ (which is the case studied in this paper) one can consider the interval between the UV boundary and the singularity at $y_s=1/\rho$, and the theory spectrum is a continuum with a mass gap of order $\rho$. The string scale is at an intermediate value, $M_s\sim 10^{-5} M_{\rm Pl}$, and the string coupling is small but larger than for the previous case, $g_s\sim 10^{-5}$. In this theory the SM cannot propagate in the bulk of the extra dimension, so we have introduced a brane, the IR or SM brane, where the SM fields are localized. We dub this theory the continuum linear dilaton model. In this theory the hierarchy problem is partly solved by the string theory, which should provide the hierarchy between $M_{\rm Pl}$ and $M_s$, i.e.~the explanation of the size of $g_s$, and by the warped extra dimensional theory, which explains the hierarchy between $M_s$  and the TeV scale.

We have considered the graviton and radion sectors in the bulk of the continuum linear dilaton model. We have worked out the general Green's functions and spectral functions for both. For the graviton, the spectrum is a continuum of KK modes with a mass gap equal to $m_g=3\rho/2$, and an isolated massless pole which corresponds to the 4D graviton. For the radion the spectrum is also a continuum of KK modes with the mass gap equal to $m_g$, and an isolated pole with a mass below $m_g$ and whose value is controlled by the Goldberger-Wise potential in the UV brane, which determines the total length of the interval. Integrating out the continuum of states, for both the graviton and radion fields, gives rise to dimension eight operators which should contribute to the effective field theory below the mass gap. These dimension eight operators give rise to anomalous quartic gauge couplings, which contribute to the low energy electroweak observables, and to high energy unitarity violations in, e.g., longitudinal vector boson scattering amplitudes. There is a wide literature on the subject and the results in this models can (and have been) easily adapted to them. 

Moreover the model can have, depending on the UV brane potential parameters, a light radion/dilaton in which case, after integration of the continuum of KK modes, it remains as the only light state on top of the SM ones. The wave function of the radion is strongly localized toward the IR so one expects its couplings with the SM fields could be sizable.
In this respect this is similar to conventional Randall-Sundrum models, so that the phenomenology of such state is expected to follow similar lines. A more phenomenological analysis of the light dilaton in the continuum linear dilaton model is beyond the scope of the present paper, and we postpone it for a future publication.

\vspace{0.5cm}
\section*{Acknowledgments}
We would like to thank O.~J.~P.~\'Eboli, M.~P\'erez-Victoria and
L.~L.~Salcedo for fruitful discussions. The work of EM is supported by
the Spanish MINEICO under Grant FIS2017-85053-C2-1-P, by the
FEDER/Junta de Andaluc\'{\i}a-Consejer\'{\i}a de Econom\'{\i}a y
Conocimiento 2014-2020 Operational Programme under Grant
A-FQM-178-UGR18, by Junta de Andaluc\'{\i}a under Grant FQM-225, and
by the Consejer\'{\i}a de Conoci\-miento, Investigaci\'on y
Universidad of the Junta de Andaluc\'{\i}a and European Regional
Development Fund (ERDF) under Grant SOMM17/6105/UGR. The research of
EM is also supported by the Ram\'on y Cajal Program of the Spanish
MINEICO under Grant RYC-2016-20678. The work of MQ is partly supported
by Spanish MINEICO under Grant FPA2017-88915-P, by the Catalan
Government under Grant 2017SGR1069, and by Severo Ochoa Excellence
Program of MINEICO under Grant SEV-2016-0588. IFAE is partially funded
by the CERCA program of the Generalitat de Catalunya.

\bibliographystyle{JHEP}
\bibliography{refs}

\providecommand{\href}[2]{#2}\begingroup\raggedright\begin{thebibliography}{10}

\bibitem{Veltman:1980mj}
M.~J.~G. Veltman, \emph{{The Infrared - Ultraviolet Connection}}, {\emph{Acta
  Phys. Polon. B} {\bfseries 12} (1981) 437}.

\bibitem{Randall:1999ee}
L.~Randall and R.~Sundrum, \emph{{A Large mass hierarchy from a small extra
  dimension}}, \href{http://dx.doi.org/10.1103/PhysRevLett.83.3370}{\emph{Phys.
  Rev. Lett.} {\bfseries 83} (1999) 3370--3373},
  [\href{https://arxiv.org/abs/hep-ph/9905221}{{\ttfamily hep-ph/9905221}}].

\bibitem{Sirunyan:2018ryr}
{\scshape CMS} collaboration, A.~M. Sirunyan et~al., \emph{{Search for resonant
  $ \mathrm{t}\overline{\mathrm{t}} $ production in proton-proton collisions at
  $ \sqrt{s}=13 $ TeV}},
  \href{http://dx.doi.org/10.1007/JHEP04(2019)031}{\emph{JHEP} {\bfseries 04}
  (2019) 031}, [\href{https://arxiv.org/abs/1810.05905}{{\ttfamily
  1810.05905}}].

\bibitem{Aaboud:2019roo}
{\scshape ATLAS} collaboration, M.~Aaboud et~al., \emph{{Search for heavy
  particles decaying into a top-quark pair in the fully hadronic final state in
  $pp$ collisions at $\sqrt{s} =$ 13 TeV with the ATLAS detector}},
  \href{http://dx.doi.org/10.1103/PhysRevD.99.092004}{\emph{Phys. Rev. D}
  {\bfseries 99} (2019) 092004},
  [\href{https://arxiv.org/abs/1902.10077}{{\ttfamily 1902.10077}}].

\bibitem{Megias:2020vek}
E.~Meg\'\i{}as, G.~Nardini and M.~Quir\'os, \emph{{Gravitational Imprints from
  Heavy Kaluza-Klein Resonances}},
  \href{http://dx.doi.org/10.1103/PhysRevD.102.055004}{\emph{Phys. Rev. D}
  {\bfseries 102} (2020) 055004},
  [\href{https://arxiv.org/abs/2005.04127}{{\ttfamily 2005.04127}}].

\bibitem{Megias:2021rgh}
E.~Meg\'\i{}as, G.~Nardini and M.~Quir\'os, \emph{{Radion dynamics, heavy
  Kaluza-Klein resonances and gravitational waves}},  in \emph{{9th
  International Conference on New Frontiers in Physics}}, 3, 2021.
\newblock \href{https://arxiv.org/abs/2103.02705}{{\ttfamily 2103.02705}}.

\bibitem{Escribano:2021jne}
R.~Escribano, M.~Mendizabal, M.~Quir\'os and E.~Royo, \emph{{On Broad
  Kaluza-Klein Gluons}},  \href{https://arxiv.org/abs/2102.11241}{{\ttfamily
  2102.11241}}.

\bibitem{Csaki:2018kxb}
C.~Csaki, G.~Lee, S.~J. Lee, S.~Lombardo and O.~Telem, \emph{{Continuum
  Naturalness}}, \href{http://dx.doi.org/10.1007/JHEP03(2019)142}{\emph{JHEP}
  {\bfseries 03} (2019) 142},
  [\href{https://arxiv.org/abs/1811.06019}{{\ttfamily 1811.06019}}].

\bibitem{Megias:2019vdb}
E.~Meg\'\i{}as and M.~Quir\'os, \emph{{Gapped Continuum Kaluza-Klein
  spectrum}}, \href{http://dx.doi.org/10.1007/JHEP08(2019)166}{\emph{JHEP}
  {\bfseries 08} (2019) 166},
  [\href{https://arxiv.org/abs/1905.07364}{{\ttfamily 1905.07364}}].

\bibitem{Goldberger:1999uk}
W.~D. Goldberger and M.~B. Wise, \emph{{Modulus stabilization with bulk
  fields}}, \href{http://dx.doi.org/10.1103/PhysRevLett.83.4922}{\emph{Phys.
  Rev. Lett.} {\bfseries 83} (1999) 4922--4925},
  [\href{https://arxiv.org/abs/hep-ph/9907447}{{\ttfamily hep-ph/9907447}}].

\bibitem{Cabrer:2009we}
J.~A. Cabrer, G.~von Gersdorff and M.~Quir\'os, \emph{{Soft-Wall
  Stabilization}},
  \href{http://dx.doi.org/10.1088/1367-2630/12/7/075012}{\emph{New J. Phys.}
  {\bfseries 12} (2010) 075012},
  [\href{https://arxiv.org/abs/0907.5361}{{\ttfamily 0907.5361}}].

\bibitem{Aharony:1999ks}
O.~Aharony, \emph{{A Brief review of 'little string theories'}},
  \href{http://dx.doi.org/10.1088/0264-9381/17/5/302}{\emph{Class. Quant.
  Grav.} {\bfseries 17} (2000) 929--938},
  [\href{https://arxiv.org/abs/hep-th/9911147}{{\ttfamily hep-th/9911147}}].

\bibitem{Antoniadis:2011qw}
I.~Antoniadis, A.~Arvanitaki, S.~Dimopoulos and A.~Giveon, \emph{{Phenomenology
  of TeV Little String Theory from Holography}},
  \href{http://dx.doi.org/10.1103/PhysRevLett.108.081602}{\emph{Phys. Rev.
  Lett.} {\bfseries 108} (2012) 081602},
  [\href{https://arxiv.org/abs/1102.4043}{{\ttfamily 1102.4043}}].

\bibitem{Cox:2012ee}
P.~Cox and T.~Gherghetta, \emph{{Radion Dynamics and Phenomenology in the
  Linear Dilaton Model}},
  \href{http://dx.doi.org/10.1007/JHEP05(2012)149}{\emph{JHEP} {\bfseries 05}
  (2012) 149}, [\href{https://arxiv.org/abs/1203.5870}{{\ttfamily 1203.5870}}].

\bibitem{Giudice:2017fmj}
G.~F. Giudice, Y.~Kats, M.~McCullough, R.~Torre and A.~Urbano,
  \emph{{Clockwork/linear dilaton: structure and phenomenology}},
  \href{http://dx.doi.org/10.1007/JHEP06(2018)009}{\emph{JHEP} {\bfseries 06}
  (2018) 009}, [\href{https://arxiv.org/abs/1711.08437}{{\ttfamily
  1711.08437}}].

\bibitem{Davoudiasl:1999jd}
H.~Davoudiasl, J.~L. Hewett and T.~G. Rizzo, \emph{{Phenomenology of the
  Randall-Sundrum Gauge Hierarchy Model}},
  \href{http://dx.doi.org/10.1103/PhysRevLett.84.2080}{\emph{Phys. Rev. Lett.}
  {\bfseries 84} (2000) 2080},
  [\href{https://arxiv.org/abs/hep-ph/9909255}{{\ttfamily hep-ph/9909255}}].

\bibitem{Davoudiasl:2001uj}
H.~Davoudiasl and T.~G. Rizzo, \emph{{Bulk physics at a graviton factory}},
  \href{http://dx.doi.org/10.1016/S0370-2693(01)00704-3}{\emph{Phys. Lett. B}
  {\bfseries 512} (2001) 100--106},
  [\href{https://arxiv.org/abs/hep-ph/0104199}{{\ttfamily hep-ph/0104199}}].

\bibitem{Fitzpatrick:2007qr}
A.~L. Fitzpatrick, J.~Kaplan, L.~Randall and L.-T. Wang, \emph{{Searching for
  the Kaluza-Klein Graviton in Bulk RS Models}},
  \href{http://dx.doi.org/10.1088/1126-6708/2007/09/013}{\emph{JHEP} {\bfseries
  09} (2007) 013}, [\href{https://arxiv.org/abs/hep-ph/0701150}{{\ttfamily
  hep-ph/0701150}}].

\bibitem{Agashe:2007zd}
K.~Agashe, H.~Davoudiasl, G.~Perez and A.~Soni, \emph{{Warped Gravitons at the
  LHC and Beyond}},
  \href{http://dx.doi.org/10.1103/PhysRevD.76.036006}{\emph{Phys. Rev. D}
  {\bfseries 76} (2007) 036006},
  [\href{https://arxiv.org/abs/hep-ph/0701186}{{\ttfamily hep-ph/0701186}}].

\bibitem{Antipin:2007pi}
O.~Antipin, D.~Atwood and A.~Soni, \emph{{Search for RS gravitons via W(L)W(L)
  decays}}, \href{http://dx.doi.org/10.1016/j.physletb.2008.07.009}{\emph{Phys.
  Lett. B} {\bfseries 666} (2008) 155--161},
  [\href{https://arxiv.org/abs/0711.3175}{{\ttfamily 0711.3175}}].

\bibitem{Dillon:2016fgw}
B.~M. Dillon and V.~Sanz, \emph{{Kaluza-Klein gravitons at LHC2}},
  \href{http://dx.doi.org/10.1103/PhysRevD.96.035008}{\emph{Phys. Rev. D}
  {\bfseries 96} (2017) 035008},
  [\href{https://arxiv.org/abs/1603.09550}{{\ttfamily 1603.09550}}].

\bibitem{Gubser:2000nd}
S.~S. Gubser, \emph{{Curvature singularities: The Good, the bad, and the
  naked}}, \href{http://dx.doi.org/10.4310/ATMP.2000.v4.n3.a6}{\emph{Adv.
  Theor. Math. Phys.} {\bfseries 4} (2000) 679--745},
  [\href{https://arxiv.org/abs/hep-th/0002160}{{\ttfamily hep-th/0002160}}].

\bibitem{York:1972sj}
J.~W. York, Jr., \emph{{Role of conformal three geometry in the dynamics of
  gravitation}},
  \href{http://dx.doi.org/10.1103/PhysRevLett.28.1082}{\emph{Phys. Rev. Lett.}
  {\bfseries 28} (1972) 1082--1085}.

\bibitem{Gibbons:1976ue}
G.~W. Gibbons and S.~W. Hawking, \emph{{Action Integrals and Partition
  Functions in Quantum Gravity}},
  \href{http://dx.doi.org/10.1103/PhysRevD.15.2752}{\emph{Phys. Rev.}
  {\bfseries D15} (1977) 2752--2756}.

\bibitem{Megias:2018sxv}
E.~Meg\'\i{}as, G.~Nardini and M.~Quir\'os, \emph{{Cosmological Phase
  Transitions in Warped Space: Gravitational Waves and Collider Signatures}},
  \href{http://dx.doi.org/10.1007/JHEP09(2018)095}{\emph{JHEP} {\bfseries 09}
  (2018) 095}, [\href{https://arxiv.org/abs/1806.04877}{{\ttfamily
  1806.04877}}].

\bibitem{DeWolfe:1999cp}
O.~DeWolfe, D.~Z. Freedman, S.~S. Gubser and A.~Karch, \emph{{Modeling the
  fifth-dimension with scalars and gravity}},
  \href{http://dx.doi.org/10.1103/PhysRevD.62.046008}{\emph{Phys. Rev.}
  {\bfseries D62} (2000) 046008},
  [\href{https://arxiv.org/abs/hep-th/9909134}{{\ttfamily hep-th/9909134}}].

\bibitem{Gubser:1999vj}
S.~S. Gubser, \emph{{AdS / CFT and gravity}},
  \href{http://dx.doi.org/10.1103/PhysRevD.63.084017}{\emph{Phys. Rev.}
  {\bfseries D63} (2001) 084017},
  [\href{https://arxiv.org/abs/hep-th/9912001}{{\ttfamily hep-th/9912001}}].

\bibitem{ArkaniHamed:1998rs}
N.~Arkani-Hamed, S.~Dimopoulos and G.~R. Dvali, \emph{{The Hierarchy problem
  and new dimensions at a millimeter}},
  \href{http://dx.doi.org/10.1016/S0370-2693(98)00466-3}{\emph{Phys. Lett. B}
  {\bfseries 429} (1998) 263--272},
  [\href{https://arxiv.org/abs/hep-ph/9803315}{{\ttfamily hep-ph/9803315}}].

\bibitem{Antoniadis:2001sw}
I.~Antoniadis, S.~Dimopoulos and A.~Giveon, \emph{{Little string theory at a
  TeV}}, \href{http://dx.doi.org/10.1088/1126-6708/2001/05/055}{\emph{JHEP}
  {\bfseries 05} (2001) 055},
  [\href{https://arxiv.org/abs/hep-th/0103033}{{\ttfamily hep-th/0103033}}].

\bibitem{Hagiwara:2008jb}
K.~Hagiwara, J.~Kanzaki, Q.~Li and K.~Mawatari, \emph{{HELAS and
  MadGraph/MadEvent with spin-2 particles}},
  \href{http://dx.doi.org/10.1140/epjc/s10052-008-0663-x}{\emph{Eur. Phys. J.
  C} {\bfseries 56} (2008) 435--447},
  [\href{https://arxiv.org/abs/0805.2554}{{\ttfamily 0805.2554}}].

\bibitem{Costantino:2020vdu}
A.~Costantino and S.~Fichet, \emph{{Opacity from Loops in AdS}},
  \href{http://dx.doi.org/10.1007/JHEP02(2021)089}{\emph{JHEP} {\bfseries 02}
  (2021) 089}, [\href{https://arxiv.org/abs/2011.06603}{{\ttfamily
  2011.06603}}].

\bibitem{Georgi:2007ek}
H.~Georgi, \emph{{Unparticle physics}},
  \href{http://dx.doi.org/10.1103/PhysRevLett.98.221601}{\emph{Phys. Rev.
  Lett.} {\bfseries 98} (2007) 221601},
  [\href{https://arxiv.org/abs/hep-ph/0703260}{{\ttfamily hep-ph/0703260}}].

\bibitem{Delgado:2008gj}
A.~Delgado, J.~R. Espinosa, J.~M. No and M.~Quir\'os, \emph{{A Note on
  Unparticle Decays}},
  \href{http://dx.doi.org/10.1103/PhysRevD.79.055011}{\emph{Phys. Rev.}
  {\bfseries D79} (2009) 055011},
  [\href{https://arxiv.org/abs/0812.1170}{{\ttfamily 0812.1170}}].

\bibitem{Eboli:2006wa}
O.~J.~P. \'Eboli, M.~C. Gonzalez-Garcia and J.~K. Mizukoshi, \emph{{p p
  ---\ensuremath{>} j j e+- mu+- nu nu and j j e+- mu-+ nu nu at O(
  alpha(em)**6) and O(alpha(em)**4 alpha(s)**2) for the study of the quartic
  electroweak gauge boson vertex at CERN LHC}},
  \href{http://dx.doi.org/10.1103/PhysRevD.74.073005}{\emph{Phys. Rev. D}
  {\bfseries 74} (2006) 073005},
  [\href{https://arxiv.org/abs/hep-ph/0606118}{{\ttfamily hep-ph/0606118}}].

\bibitem{Almeida:2020ylr}
E.~d.~S. Almeida, O.~J.~P. \'Eboli and M.~C. Gonzalez\textendash{}Garcia,
  \emph{{Unitarity constraints on anomalous quartic couplings}},
  \href{http://dx.doi.org/10.1103/PhysRevD.101.113003}{\emph{Phys. Rev. D}
  {\bfseries 101} (2020) 113003},
  [\href{https://arxiv.org/abs/2004.05174}{{\ttfamily 2004.05174}}].

\bibitem{Sirunyan:2019der}
{\scshape CMS} collaboration, A.~M. Sirunyan et~al., \emph{{Search for
  anomalous electroweak production of vector boson pairs in association with
  two jets in proton-proton collisions at 13 TeV}},
  \href{http://dx.doi.org/10.1016/j.physletb.2019.134985}{\emph{Phys. Lett. B}
  {\bfseries 798} (2019) 134985},
  [\href{https://arxiv.org/abs/1905.07445}{{\ttfamily 1905.07445}}].

\bibitem{Sirunyan:2020tlu}
{\scshape CMS} collaboration, A.~M. Sirunyan et~al., \emph{{Measurement of the
  cross section for electroweak production of a Z boson, a photon and two jets
  in proton-proton collisions at $\sqrt{s} =$ 13 TeV and constraints on
  anomalous quartic couplings}},
  \href{http://dx.doi.org/10.1007/JHEP06(2020)076}{\emph{JHEP} {\bfseries 06}
  (2020) 076}, [\href{https://arxiv.org/abs/2002.09902}{{\ttfamily
  2002.09902}}].

\bibitem{Sirunyan:2020gyx}
{\scshape CMS} collaboration, A.~M. Sirunyan et~al., \emph{{Measurements of
  production cross sections of WZ and same-sign WW boson pairs in association
  with two jets in proton-proton collisions at $\sqrt{s} =$ 13 TeV}},
  \href{http://dx.doi.org/10.1016/j.physletb.2020.135710}{\emph{Phys. Lett. B}
  {\bfseries 809} (2020) 135710},
  [\href{https://arxiv.org/abs/2005.01173}{{\ttfamily 2005.01173}}].

\bibitem{Ari:2021ixv}
V.~Ari, E.~Gurkanli, M.~K\"oksal, A.~Guti\'errez-Rodr\'\i{}guez and M.~A.
  Hern\'andez-Ru\'\i{}z, \emph{{Future projections on the anomalous
  $WW\gamma\gamma$ couplings in hadron-hadron interactions at the FCC-hh}},
  \href{https://arxiv.org/abs/2104.00474}{{\ttfamily 2104.00474}}.

\bibitem{Guo:2020lim}
Y.-C. Guo, Y.-Y. Wang, J.-C. Yang and C.-X. Yue, \emph{{Constraints on
  anomalous quartic gauge couplings via $W\gamma jj$ production at the LHC}},
  \href{http://dx.doi.org/10.1088/1674-1137/abb4d2}{\emph{Chin. Phys. C}
  {\bfseries 44} (2020) 123105},
  [\href{https://arxiv.org/abs/2002.03326}{{\ttfamily 2002.03326}}].

\bibitem{Csaki:2000zn}
Csaki, M.~L. Graesser and G.~D. Kribs, \emph{{Radion dynamics and electroweak
  physics}}, \href{http://dx.doi.org/10.1103/PhysRevD.63.065002}{\emph{Phys.
  Rev.} {\bfseries D63} (2001) 065002},
  [\href{https://arxiv.org/abs/hep-th/0008151}{{\ttfamily hep-th/0008151}}].

\bibitem{Megias:2015ory}
E.~Meg\'\i{}as, O.~Pujol\`as and M.~Quir\'os, \emph{{On dilatons and the LHC
  diphoton excess}},
  \href{http://dx.doi.org/10.1007/JHEP05(2016)137}{\emph{JHEP} {\bfseries 05}
  (2016) 137}, [\href{https://arxiv.org/abs/1512.06106}{{\ttfamily
  1512.06106}}].

\end{thebibliography}\endgroup

\end{document}